\documentclass[journal=jpcbfk,manuscript=article,layout=traditional]{achemso}
\usepackage[version=3]{mhchem} 
\usepackage[T1]{fontenc}      
\usepackage{graphicx}
\usepackage{subfigure}
\usepackage{multirow}
\usepackage{hhline}
\usepackage{chemarr}
\usepackage{amssymb}
\usepackage{color}
\usepackage{xspace}
\usepackage{chngcntr}
\counterwithout{table}{section}
\usepackage{textcomp}
\usepackage{gensymb}
\usepackage{xr} 
\usepackage{pdfpages}
\usepackage{booktabs}
\usepackage{multirow}
\usepackage[table,xcdraw]{xcolor}
\usepackage{colortbl} 
\usepackage{bm}
\usepackage{float}
\usepackage{ulem}

\usepackage{float}
\usepackage{graphicx} 
\usepackage{hyperref}
\usepackage{cleveref}
\usepackage{soul}
\usepackage{xcolor}
\usepackage{caption}
\usepackage{float}
\usepackage{float}
\usepackage{siunitx}
\usepackage{lineno} 

\usepackage{marginnote}
\usepackage{enumitem}

\newcommand{\angstrom}{\mbox{\normalfont\AA}}

\author{Narjes Ansari\textsuperscript{$\nabla$}}\affiliation{Qubit Pharmaceuticals, Advanced Research Department, 75014 Paris, France}\email{narjesa@qubit-pharmaceuticals.com}
\author{Félix Aviat\textsuperscript{$\nabla$}}\affiliation{Qubit Pharmaceuticals, Advanced Research Department, 75014 Paris, France}\email{felix.aviat@qubit-pharmaceuticals.com}
\author{Jérôme Hénin}\affiliation{Laboratoire de Biochimie Théorique,  UPR 9080 CNRS, Université de Paris Cité, 75005 Paris, France}
\author{Jean-Philip Piquemal}\affiliation{Qubit Pharmaceuticals, Advanced Research Department, 75014 Paris, France}\altaffiliation{Laboratoire de Chimie Théorique, Sorbonne Université, UMR 7616 CNRS, 75005 Paris, France}
\author{Louis Lagardère}\affiliation{Qubit Pharmaceuticals, Advanced Research Department, 75014 Paris, France}\altaffiliation{Laboratoire de Chimie Théorique, Sorbonne Université, UMR 7616 CNRS, 75005 Paris, France}\email{louis.lagardere@sorbonne-universite.fr}

\title{$\text{Dual-LAO}$ for calculating fast and robust relative binding free energies of simple and complex transformations}

\begin{document}

\begin{center}
\textsuperscript{$\nabla$} N.A. and F.A. contributed equally.   
\end{center}

\maketitle

\begin{abstract}
Relative Binding Free Energy ($\text{RBFE}$) calculations are a cornerstone of rational hit-to-lead and lead optimization in modern drug discovery. However, the high computational cost and limited reliability in tackling large or complex molecular transformations often prevent their routine, high-throughput use.
Here we introduce $\text{Dual-LAO}$, a novel, highly efficient method for calculating $\text{RBFE}$. Building on the $\text{Lambda-ABF-OPES}$ framework, this method combines a dual-topology setup and suitable restraints to dramatically accelerate free energy convergence.
We demonstrate that $\text{Dual-LAO}$, in combination with the AMOEBA polarizable force field, achieves an unprecedented acceleration factor of 15 to 30 times compared to current state-of-the-art methods on standard drug targets. Crucially, the approach maintains high accuracy and successfully tackles previously prohibitive molecular changes, including scaffold-hopping, buried water displacement, charge changes, ring-opening, and binding pose perturbations.
This significant leap in efficiency allows for the widespread, routine integration of predictive molecular simulations into the rapid optimization cycles of drug discovery, enabling chemists to confidently model historically challenging systems in timescales compatible with real-world project deadlines.

\end{abstract}

\section{Introduction}

Accelerating therapeutic development critically hinges on accurately predicting the binding affinity ($\Delta G$) between small molecules and macromolecular targets~\cite{de2008computational}. Computational methods play a pivotal role, offering the potential to streamline the costly and time-consuming process of drug discovery by prioritizing promising candidates for experimental investigation~\cite{jorgensen2009cadd, homeyer2014, williams2018free}. Among the most rigorous computational tools are physics-based alchemical free energy (AFE) methods, which leverage statistical mechanics and molecular simulations~\cite{CHODERA2011150}. Alchemical methods rely on the exploration of non-physical intermediate states to perform a transformation between two physical end-states.

One class of such methods, Absolute Binding Free Energy ($\text{ABFE}$) calculations, aims to compute the total free energy change upon ligand binding~\cite{Merz2020EvolutionOfAFE}. While conceptually appealing, ABFE prediction remains computationally demanding, often struggling with slow convergence due to the extensive conformational sampling required for both the binding process and the large alchemical transformations involved (e.g., full ligand decoupling)~\cite{GALLICCHIO2011161}. Achieving reliable $\text{ABFE}$ results necessitates both accurate force fields capable of capturing subtle molecular interactions (including polarization)~\cite{amezcua_sampl7_2021} and highly efficient sampling techniques to overcome significant free energy barriers~\cite{henin2022enhanced,torrie1977nonphysical,darve2001calculating,darve2008adaptive,comer2015adaptive,laio2002escaping,barducci2011metadynamics,Invernizzi2020,Invernizzi2022,maragliano2006temperature}.

Consequently, $\text{RBFE}$ calculations have become the predominant approach in hit-to-lead and lead optimization campaigns~\cite{cournia_relative_2017}. $\text{RBFE}$ methods calculate the difference in binding affinity ($\Delta\Delta G$) between two structurally related ligands ($L_1$ and $L_2$) via a thermodynamic cycle\cite{pearlman_overlooked_1991, york_2023_modern} by alchemically transforming $L_1$ into $L_2$ in both the protein-bound state ($\Delta G_{\text{complex}}$) and the unbound, solvated state ($\Delta G_{\text{solv}}$):
\begin{equation}
\Delta\Delta G = \Delta G_{\text{complex}} - \Delta G_{\text{solv}} 
\label{eq:ddgcycle}
\end{equation}
These methods benefit from significant error cancellation and typically involve smaller, more manageable perturbations than $\text{ABFEs}$~\cite{cournia_relative_2017}. This makes $\text{RBFE}$ less computationally intensive than $\text{ABFE}$ and often capable of higher precision for comparing congeneric series~\cite{cournia_relative_2017}. Furthermore, the computational cost for all these simulations has been dramatically reduced by the use of Graphics Processing Units (GPUs), enabling $\text{RBFE}$ calculations to be performed with sufficient throughput to impact active drug discovery projects ~\cite{cournia_relative_2017}.

$\text{RBFE}$ are typically performed using molecular dynamics (MD). The free energy difference is commonly estimated using methods like Free Energy Perturbation (FEP)~\cite{zwanzig1954high} or Thermodynamic Integration (TI)~\cite{straatsma1991multiconfiguration}, often involving multiple simulations at discrete intermediate states along the transformation path. 

To improve efficiency, alternative approaches have been developed. These include nonequilibrium methods~\cite{gapsys2012Nonequilibrium}, which are highly parallelizable but may require significant computational effort to reach the same level of precision as equilibrium methods, a trade-off that remains a subject of ongoing discussion in the community~\cite{kalpokas2025accurate}. Another class of methods involves expanded ensemble techniques, where the alchemical coupling parameter $\lambda$ is treated as an extended variable that evolves during a single simulation. Prominent examples include $\lambda$-dynamics ($\lambda$D)~\cite{kong1996lambda} and its extension, multisite $\lambda$-dynamics (MS$\lambda$D)~\cite{brooks2011MultisiteLdyn}. MS$\lambda$D is particularly scalable, allowing simultaneous perturbations at multiple molecular sites and enabling the exploration of large combinatorial chemical spaces with significantly fewer simulations than traditional FEP or TI.

To practically model this alchemical transformation, several topology schemes can be employed. The \textit{single-topology} approach requires mapping the atoms of $L_1$ onto $L_2$ and introducing ``dummy'' atoms for non-matching regions, which can become complex for dissimilar ligands~\cite{jorgensen_monte_1985,pearlman_overlooked_1991}. The \textit{dual-topology} approach simulates both $L_1$ and $L_2$ simultaneously, with the interactions of one ligand being ``turned off'' while the other is ``turned on''~\cite{gao_hidden_1989,pearlman_overlooked_1991}. This avoids dummy atoms but can require careful spatial restraints to manage the non-interacting components~\cite{ries_restraintmaker_2022}. An alternative is the \textit{hybrid} or \textit{modified dual-topology} scheme, which defines a maximal common substructure (MCS) as a single, shared entity while treating the non-shared ``appearing'' and ``disappearing'' regions separately, thus avoiding both dummy atoms and the need for complex restraints~\cite{ERIKSSON1995hybridtop_anonymous,Wei2019hybridtop}.

Modern $\text{RBFE}$ protocols, combining advanced force fields (e.g., OPLS3e~\cite{harder2016OPLS3}), enhanced sampling techniques (e.g., REST2~\cite{Wang2011REST2}), and automated workflows~\cite{cournia_relative_2017}, have demonstrated impressive performance in numerous retrospective and prospective studies, frequently reporting root-mean-square errors (RMSE) relative to experiment approaching 1 kcal.mol\textsuperscript{-1}~\cite{wang2015accurate,ross2023maximal,azimi2022relative}. This level of accuracy, often designated as ``chemical accuracy'', allows $\text{RBFE}$ predictions to effectively rank compounds and guide synthesis prioritization~\cite{wang2015accurate}. Large-scale prospective studies have confirmed the value of $\text{RBFE}$ in enriching hit rates for potent compounds compared to other methods~\cite{schindler2020large}.

Despite these successes, $\text{RBFE}$ methods face significant challenges that can impact their reliability and accuracy\cite{york2023modern,CHODERA2011150}. Accuracy is intrinsically limited by the underlying force field and, critically, by the extent of conformational sampling ~\cite{muegge2023recent}. Insufficient sampling is a frequent pitfall, exacerbated by the chaotic nature of molecular dynamics, which mandates the use of ensemble simulations for statistically robust and reproducible results~\cite{pohorille2010good}. Furthermore, the resulting free energy distributions are often non-Gaussian, complicating error analysis~\cite{khuttan2024make}. Several specific scenarios are notoriously difficult: transformations involving large structural rearrangements (macrocyclization or scaffold hopping~\cite{ghidini2025bidirectional,paggi2021leveraging}), changes in ligand net charge~\cite{rocklin2013blind}, covalent inhibitors~\cite{cournia_relative_2017}, uncertainties in the initial binding pose~\cite{paggi2021leveraging}, and perturbations involving the displacement or reorganization of buried water molecules~\cite{chodera2011alchemical}. Ambiguities in protein structure (e.g., missing loops) also pose challenges, often necessitating retrospective validation before prospective use. Furthermore, uncertainties in the precise binding site configuration and the ligand’s initial binding pose represent the primary bottlenecks for the prospective application of BFE calculations~\cite{behera2025quantification}. It is also crucial to contextualize computational accuracy targets against experimental reproducibility limits; independent experimental measurements of the same affinity can differ by 0.8 -- 1.0 kcal.mol\textsuperscript{-1} (RMSE $\approx 0.6 \; \text{to}\; 0.7~pK_i$ units), setting a practical limit on achievable computational accuracy ~\cite{kramer2012experimental}.

Recently, the $\text{Lambda-ABF-OPES}$ method was introduced as a highly efficient hybrid approach for calculating ABFE~\cite{ansari2025lambda}. By coupling Lambda-adaptive biasing force (Lambda-ABF) with On-the-fly Probability Enhanced Sampling (OPES), this technique achieves significantly faster convergence --- up to 9-fold improvement compared to the original Lambda-ABF --- while maintaining high accuracy, demonstrated with a mean absolute error of approximately 0.9 kcal.mol\textsuperscript{-1} for challenging bromodomain inhibitors \cite{ansari2025lambda}.

Lambda-ABF-OPES sampling acceleration also relies on the use of several simulations running simultaneously, called ``walkers'', exchanging at fixed intervals the information gathered by their respective ABF-estimators. Each such walker thus benefits from the exploration of the others, which effectively increases the rate of exploration for all.

A key aspect contributing to the accuracy reached in this work is the use of the advanced AMOEBA polarizable force field~\cite{ansari2025lambda}. Given the established difficulties in sampling for both $\text{ABFE}$ and $\text{RBFE}$, applying such a demonstrably rapid and accurate methodology developed for ABFE within the $\text{RBFE}$ context holds significant promise.

In this work, we introduce $\text{Dual-LAO}$ (Dual-topology, Dual-DBC Lambda-ABF-OPES), a novel $\text{RBFE}$ strategy designed to significantly accelerate convergence and improve accuracy. This method integrates three core components: a dual-topology representation of the transformed ligands, enhanced alchemical sampling using the $\text{Lambda-ABF-OPES}$~\cite{ansari2025lambda} framework, and a dual-DBC scheme utilizing specific distance-to-bound configuration ($\text{DBC}$) restraints. We implement this approach within the Tinker-HP simulation package, leveraging the advanced $\text{AMOEBA}$ polarizable force field \cite{adjoua_tinker-hp_2021}. This combined approach benefits from the sampling power of $\text{Lambda-ABF-OPES}$, the flexibility of the dual topology method, and narrowing down of phase space volume due to the DBC restraints.

In the following, we rigorously evaluate the performance of this implementation across a spectrum of challenging scenarios frequently encountered in drug discovery. These transformations cover four distinct types, each posing unique challenges to sampling and convergence: (i) fragment-based transformations, including difficult charge-changing perturbations 
, (ii) perturbations involving problematic buried water displacement
, (iii) fragment-like chemical perturbations
, and (iv) complex scaffold-hopping perturbations, featuring challenging ring-opening transformations
.

Our findings demonstrate the high accuracy and robustness of the $\text{Dual-LAO}$ method across this diverse and challenging set. Across the entire dataset comprising 30 compounds and 56 alchemical edges, the approach yields an overall edgewise $\text{RMSE}$ of 0.52~kcal.mol\textsuperscript{-1} and a pairwise $\text{RMSE}$ of 0.56 kcal.mol\textsuperscript{-1}, with a coefficient of determination ($\text{R}^{2}$) of 0.90. Crucially, this high level of accuracy is achieved with a simulation time of only 4~ns/walker for the complex and 2~ns/walker for the solvent phase per transformation (using 4 walkers). These results demonstrate that the $\text{Dual-LAO}$ approach consistently achieves rapid convergence and high accuracy, with $\text{RMSE}$ values below 1 kcal.mol\textsuperscript{-1} across all these $\text{RBFE}$ test systems, in significantly smaller simulation time than conventional $\text{RBFE}$ methods.

To further accelerate calculations, particularly when assessing multiple related analogues, we implemented Multiple topology $\text{Dual-LAO}$ within the Tinker-HP package. This extension allows for the simultaneous sampling of multiple alchemical transformations within a single simulation framework, leveraging the efficiency benefits observed in other multiple topology dynamic lambda approaches~\cite{hayes2021blade}. In this framework, all free energy differences are obtained concurrently from just one simulation run. As validation of this implementation, we calculated the relative hydration free energies for 15 distinct molecule pairs within the benzene family.


\section{Results and Discussion}
This section details the $\text{RBFE}$ results obtained from various alchemical transformations using  $\text{Dual-LAO}$ method. These transformations cover: \text{fragment-based transformations}, 
\text{buried water displacement}, 
\text{fragment-like chemical perturbations}, 
and \text{scaffold-hopping perturbations}. 
Representative examples are provided for each type.

\subsection{$\text{RBFEs}$ of Fragments for the PWWP1 Domain of NSD3} \label{sec:PWWP1}

\begin{figure}[H]
	\centering
	\includegraphics[width=0.8\textwidth]{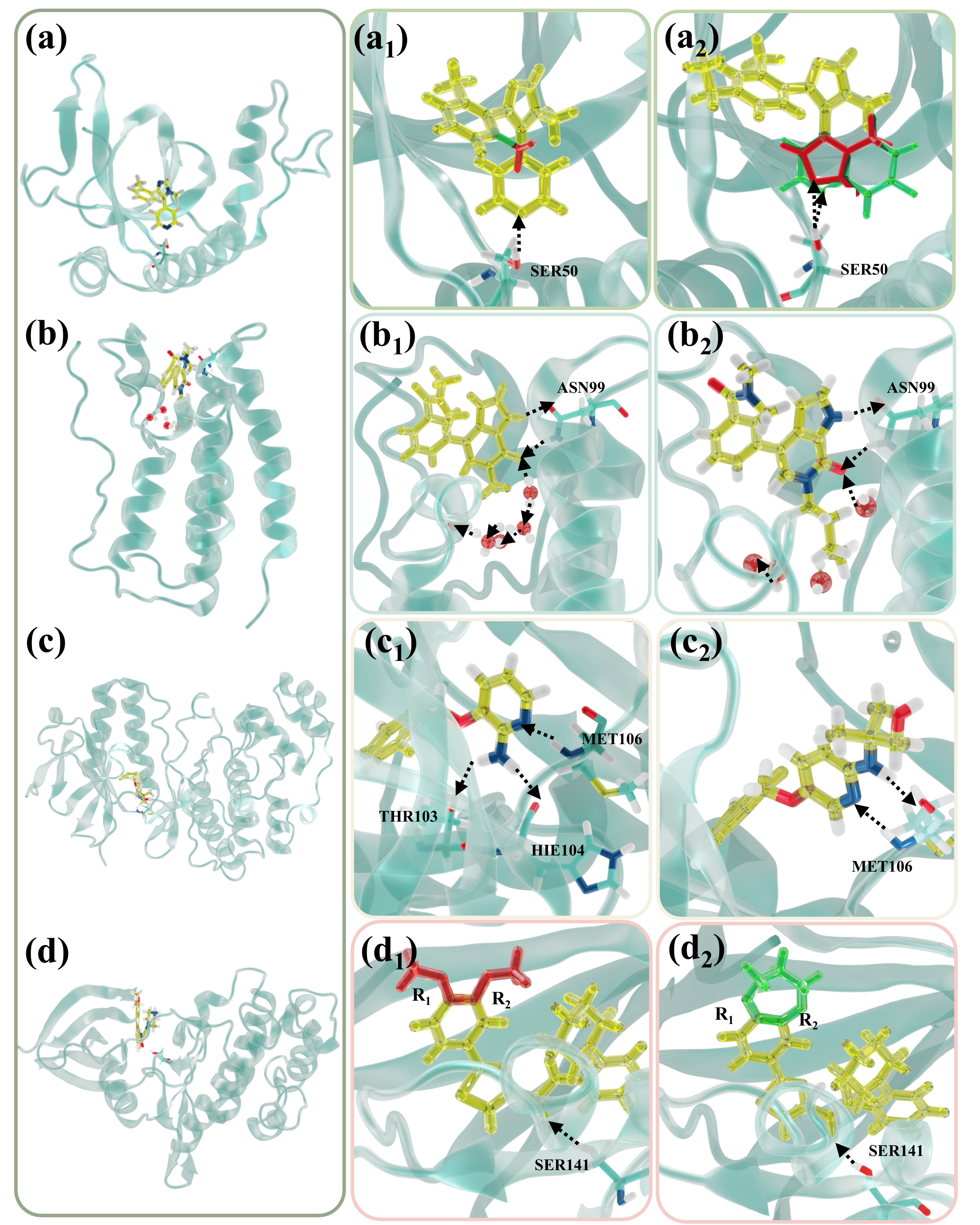}   
     \caption{\textbf{RBFE Alchemical Transformations and Systems.} (a) PWWP1-ligand complex: ($\text{a}_{1}$) shows a side chain transformation where the change involves only the peripheral substituent, keeping the core binding pose conserved. ($\text{a}_{2}$) illustrates a binding pose transformation, where Ligand A (green) adopts a different pose compared to Ligand B (red), leading to new protein-ligand interactions. (b) BRD4-ligand complex: ($\text{b}_{1}$) and ($\text{b}_{2}$) display Buried Water Displacement. The comparison shows the displacement of ordered buried water molecules (red and white spheres) as the ligand size increases and extends its binding pose. (c) P38-ligand complex: ($\text{c}_{1}$) and ($\text{c}_{2}$) represent Fragment-Based Perturbations, where the transformation causes a change in the ligand's interaction with the protein. (d) CHK1-ligand complex: ($\text{d}_{1}$) and ($\text{d}_{2}$) show Scaffold-Hopping Transformations, where the $\text{R}_{1}$ and $\text{R}_{2}$ side chains change from an open form to a closed ring structure.}
	\label{fig:diff_systems}
\end{figure}

Our relative binding free energy $\text{RBFE}$ simulations were performed as part of a fragment elaboration study targeting the PWWP1 domain of NSD3~\cite{bottcher2019fragment}. A set of $12$ ligands (2D structures of the ligands are detailed in Supplementary Fig.~S1) was strategically selected~\cite{alibay2022evaluating}, covering a significant range of binding affinities, specifically from an initial 160~$\mu$M fragment hit (ligand 8) up to the highly potent 170~nM optimized lead compound, $\text{BI-9321}$. Detailed affinity data are provided in the Supplementary data 1. The alchemical free energy network for these transformations — primarily focused on $\text{R}$-group modifications — was initially constructed using the \text{LoMap} algorithm~\cite{liu2013lead} (see Supplementary Fig.~S2). This yielded a total of $15$ alchemical transformation pairs that connected the $12$ ligands. These pairs were strategically chosen to represent different types of chemical changes and were broadly categorized into two types, as illustrated in Fig.~\ref{fig:diff_systems} panels $\text{a}_{1}$ and $\text{a}_{2}$.

\begin{enumerate}
    \item \textbf{Side Chain Transformations (Fig.~\ref{fig:diff_systems} panel $\text{a}_{1}$):} These pairs feature two ligands that share the same central binding core (shown in yellow). The example of the transformation of ligand 10 to ligand 11 is shown in this figure, where the transformation occurs solely on a peripheral substituent. This type of transformation is typically less challenging to sample in \text{RBFE} calculations because the binding mode remains conserved.
    \item \textbf{Binding Pose Transformations (Fig.~\ref{fig:diff_systems} panel $\text{a}_{2}$):} These pairs represent a more challenging scenario in which the two ligands adopt different binding poses (for example ligand 17 to ligand 9466), leading to changes in protein-ligand interactions, including hydrogen bonding (HB) and van der Waals forces. This results in a binding free energy difference of around $\text{3}$-$\text{4}$ kcal.mol\textsuperscript{-1}. Such a substantial change in the protein-ligand interaction pattern and binding pose requires robust conformational sampling and presents a stringent test for the \text{RBFE} methodology.
\end{enumerate}

In addition to the pose changes, the network included a charge transfer transformation (pair: $\text{lig10} \to \text{lig12}$, see Supplementary Fig.~S1). Such transformations, in which the charge state of a ligand changes between the initial and final states, are difficult to model accurately in the computational literature\cite{ross2023maximal}. Our calculated result for this specific, challenging pair ($\Delta\Delta G_{\text{CAL}}=-1.22 \pm 0.11 \text{ kcal.mol\textsuperscript{-1}}$) shows excellent agreement with the experimental value ($\Delta\Delta G_{\text{EXP}}=-1.04 \text{ kcal.mol\textsuperscript{-1}}$). It is noteworthy that our results achieved convergence (see Supplementary Figs.~S3--S6) in less than 4 ns/walker for the complex phase and less than 2 ns/walker for the solvent phase. This contrasts sharply with state-of-the-art methods utilizing alchemical charge-changing perturbations, which typically require 24 $\lambda$ windows, with each window simulated for 20 ns~\cite{ross2023maximal}. In total, for such challenging systems, our method demonstrates a convergence speed approximately 30 times faster. To further validate these results, we performed additional simulations incorporating co-alchemical ion corrections to maintain system neutrality; the results (detailed in Section Assessment of Charge-Change Corrections of the Supplementary Information) confirm that the standard Dual-LAO treatment effectively captures the binding physics of this system without the need for explicit correction factors.

To ensure statistical rigor of the calculations, \text{RBFE} was calculated in both directions ($L_1 \to L_2$ and $L_2 \to L_1$) for each pair, resulting in a total of $30$ calculated relative free energy values ($\Delta\Delta G_{\text{CAL}}$) throughout the network (see Supplementary data 1 and Supplementary Fig.~S2).

The calculations were performed using our dual topology framework and tested with the dual-DBC restraint scheme. The raw data, including the calculated $\Delta\Delta G_{\text{CAL}}$ and its associated uncertainty for every leg, are compiled in Supplementary data 1.

\begin{figure}[H]
	\centering
	\includegraphics[width=0.9\textwidth]{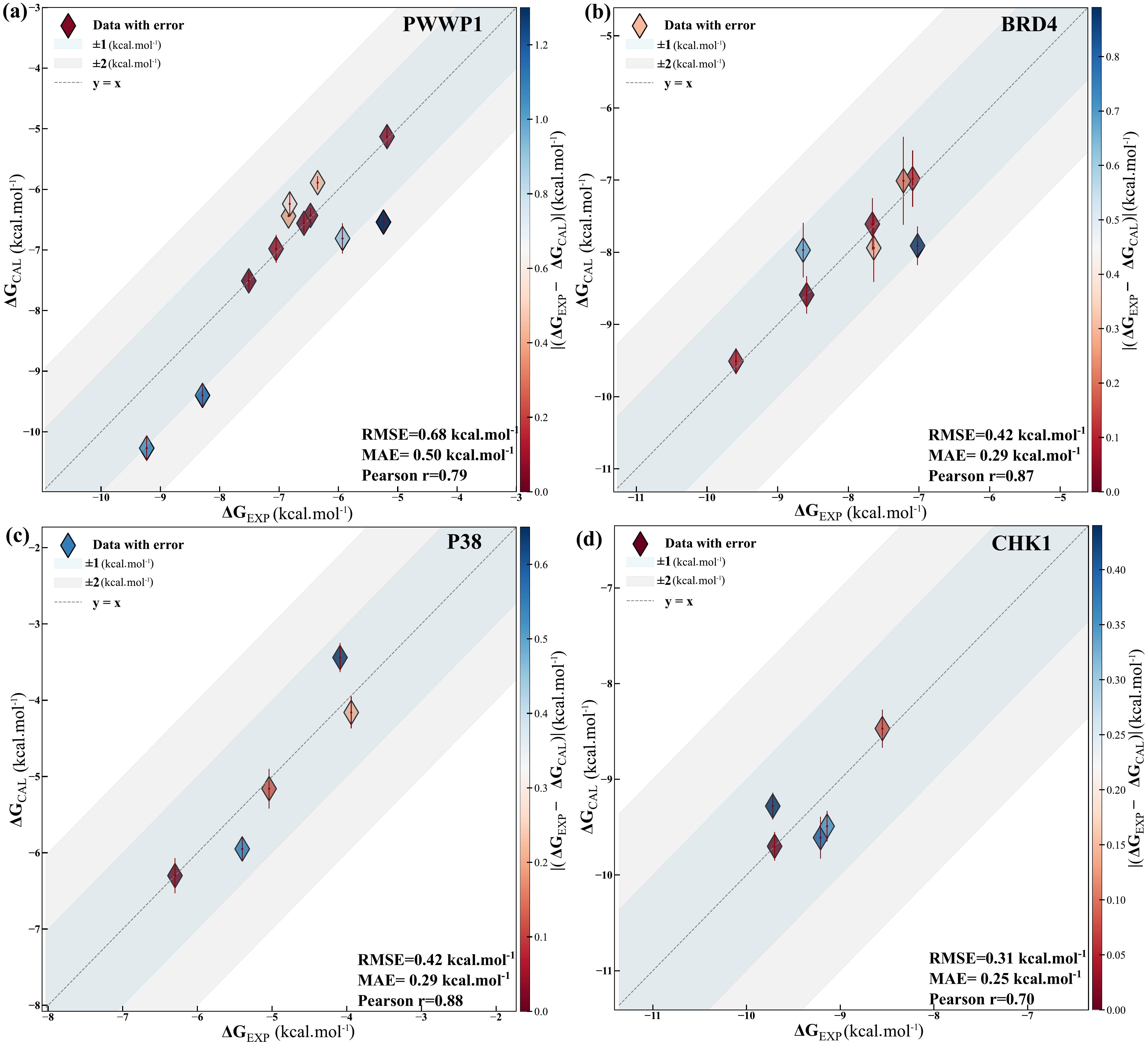}   
\caption{\textbf{Experimental vs. Calculated Absolute Binding Free Energies derived from $\text{RBFE}$.} (a) PWWP1, (b) BRD4, (c) P38, and (d) CHK1 complexes. The calculated $\Delta G_{\text{CAL}}$ results and associated errors represent the mean $\pm$ standard error derived from the \text{Weighted Least-Squares ($\text{WLS}$) fitting} of the $\text{RBFE}$ network (see SI for more details). The dark and light shaded regions indicate $\pm$1~kcal.mol\textsuperscript{-1} and $\pm$2~kcal.mol\textsuperscript{-1} deviations from the experimental values, respectively. The color bar displays the absolute difference between the experimental and computed values, $|\Delta G_{\text{EXP}} - \Delta G_{\text{CAL}}|$. For each system, key performance statistics including \text{Pearson's $r$}, \text{RMSE}, and \text{MAE} are reported.}
	\label{fig:ABFE_All}
\end{figure}

The performance of the $\text{RBFE}$ calculations was assessed by comparing the simulated values (noted $\Delta\Delta G_{\text{CAL}}$) with experimentally obtained ones (noted $\Delta \Delta G_{\text{EXP}}$). Supplementary Fig.~S7 (a) illustrates the correlation plot for the $\Delta\Delta G_{\text{CAL}}$ values, and Fig.~\ref{fig:ABFE_All}(a) shows the correlation plot for the Absolute Binding Free Energy ($\Delta G_{\text{CAL}}$) values derived from the relative network, following the method described in Supplementary Information.

The dual-DBC restraint scheme yielded excellent agreement with the experimental data, with the total RMSE for $\Delta\Delta G_{\text{CAL}}$ approximately $\text{0.46}~\text{kcal.mol\textsuperscript{-1}}$ and RMSE of $\text{0.68}~\text{kcal.mol\textsuperscript{-1}}$ for $\Delta G_{\text{CAL}}$. These values compare favorably to the work of Alibay et al.~\cite{alibay2022evaluating}, who reported an RMSE of $1.14\text{ kcal.mol\textsuperscript{-1}}$.

To further assess the robustness and convergence of our protocol, we calculated the cycle-closure for all possible closed cycles within the PWWP1 dataset (see Supplementary Fig.~S1 and Supplementary Table~S1). The calculated closure values are notably low, with a maximum absolute deviation of $0.61\text{ kcal.mol\textsuperscript{-1}}$ and standard deviations consistently remaining below $1\text{ kcal.mol\textsuperscript{-1}}$. This high degree of internal consistency suggests that no inherent bias arises from the dual-DBC protocol. This demonstrates the robustness of the methodology in different types of chemical transformations, including the pose changes and charge transfers. 

\subsection{$\text{RBFEs}$ of BRD4 with Buried Water Displacement} \label{sec:BRD4}

One of the significant challenge in both $\text{ABFE}$ and $\text{RBFE}$ simulations arises when perturbations involve displacing buried water molecules within the protein active site.

To provide another challenging test for our current $\text{Dual-LAO}$ method, we thus selected the Bromodomain 4 ($\text{BRD4}$) system, as shown in Fig.~\ref{fig:diff_systems} panels $\text{b}_{1}$ and $\text{b}_{2}$.
This target is a well-known benchmark for computational methods, recognized to be problematic for achieving accurate binding free energies. 
In our prior studies~\cite{ansari2025lambda,blazhynska2025water} utilizing the $\text{Lambda-ABF-OPES}$ methodology, we successfully calculated the $\text{ABFE}$ for 
seven targets with different variety of ligands featuring such buried water displacement events, and all results demonstrated excellent predictive accuracy for $\Delta G_{\text{CAL}}$ in these challenging systems.
In the work by Ross et al. \cite{ross2023maximal} using $\text{FEP}+$, the $\text{BRD4}$ ligand set presented considerable difficulty, resulting in an $\text{RMSE}$ of $\text{2.05} \pm \text{0.24}~\text{kcal.mol\textsuperscript{-1}}$ and an $R^{2}$ value near zero.

We applied our $\text{Dual-LAO}$ approach to this same challenging set of ligands, which consists of eight ligands and eleven alchemical transformation pairs (see Supplementary Fig.~S8 for the 2D structures of the ligands and Supplementary  Fig.~S9 for the LoMap network illustrating the various alchemical connections.) The correlation plots of $\Delta\Delta G_{\text{CAL}}$ and $\Delta G_{\text{CAL}}$ with experimental values are presented in Supplementary Fig.~S7 (b) and Fig.~\ref{fig:ABFE_All} (b), respectively. Additionally, as an example, the convergence of PMF and the $\lambda$ exploration over time for the transformation of ligand 2 (with six buried water molecules, Fig.~\ref{fig:diff_systems} panel $b_{1}$) to ligand 8 (with only four buried water molecules, Fig.~\ref{fig:diff_systems} panel $b_{2}$) are shown in Supplementary Figs.~S11-S13. Here again, the convergence was achieved in less than $4\,\text{ns}/\text{walker}$ for the complex phase and $2\,\text{ns}/\text{walker}$ for the solvent phase.

$\text{Dual-LAO}$ results show a marked improvement over the $\text{FEP}+$ benchmark, demonstrating the method's ability to handle systems involving complex buried water perturbations. Specifically, we achieved an $\text{RMSE}$ of $\text{0.64}$ kcal.mol\textsuperscript{-1} and an $R^2$ of $\text{0.74}$ for $\Delta\Delta G_{\text{CAL}}$, and an $\text{RMSE}$ of $\text{0.42}$ kcal.mol\textsuperscript{-1} and an $R^2$ of $\text{0.76}$ for $\Delta G_{\text{CAL}}$.

In order to check the proper displacement of buried water molecules during the Dual-LAO simulations, we analyzed as an example the coordination number ($\text{CN}$) of water around the $\text{O1}$ atom (see Supplementary Fig.~S10) for the alchemical transformation pair connecting $\text{ligand 2}$ ($\lambda=0$) to $\text{ligand 5}$ ($\lambda=1$). We focused specifically on this atom to ensure clear visualization and analysis of the water perturbation.

Supplementary Fig.~S10 displays the density plot of water molecules around the $\text{O1}$ atom as a function of the alchemical coupling parameter, $\lambda$. As shown in the figure, the coordination number of water molecules around the $\text{O1}$ atom is higher when the system is closer to the $\text{ligand 2}$ state ($\lambda \approx 0$) compared to the $\text{ligand 5}$ state ($\lambda \approx 1$). This difference confirms that the $\text{Dual-LAO}$ methodology successfully drives the necessary water perturbation and displacement as the ligand transforms. The success of the \text{Dual-LAO} approach, particularly in resolving these challenging water rearrangements, can be further elucidated through a comparison with stochastic methods such as \text{Non-Equilibrium Grand Canonical Monte Carlo (NE-GCMC)} \cite{melling2023enhanced}. While \text{NE-GCMC} facilitates water insertion/deletion through discrete non-equilibrium switching moves, \text{Dual-LAO} utilizes a continuous dynamical coupling enhanced by the \text{OPES$_{e}$} and \text{ABF} algorithms. As demonstrated in our previous work \cite{ansari2025lambda,blazhynska2025water}, the integration of \text{OPES$_{e}$} effectively treats the solvent environment as an orthogonal degree of freedom. By adaptively biasing the probability distribution, \text{OPES$_{e}$} actively drives the exploration of the configurational space, facilitating the transition of water molecules in and out of buried cavities. This provides a specific advantage over \text{GCMC-based} sampling in cases where water movement is strongly coupled to protein conformational plasticity. The joint alchemical-configurational landscape allows for a mutual relaxation of the binding pocket and the solvent environment, lowering the effective free energy barriers to water reorganization without the need for predefined \text{GCMC} trial volumes or discrete switching states.

\subsection{$\text{RBFEs}$ of P38 with Fragment-Like Chemical Perturbations } \label{sec:P38}

We applied the \text{Dual-LAO} method to the P38 fragment set from Ross et al.~\cite{ross2023maximal}, which comprises $\text{5}$ ligands (See Supplementary Fig.~S14 for 2D structures of the ligands) and $\text{7}$~transformations (see Supplementary Fig.~S15 for the LoMap network illustrating the various alchemical connections). Panels $\text{c}_{1}$ and $\text{c}_{2}$ in Fig.~\ref{fig:diff_systems} show a representative transformation pair. Ligand~4 forms $\text{3}$~stable hydrogen bonds (HBs) with residues THR103, HIE104, and MET106 (see Fig.~\ref{fig:diff_systems}~$\text{c}_{1}$), while Ligand~7 interacts only with residue MET106 via $\text{2}$~HBs (see Fig.~\ref{fig:diff_systems}~$\text{c}_{2}$). In this specific perturbation, the left side of the two ligands is identical, while the right side undergoes a complex change. The convergence of PMF and the $\lambda$ exploration over time for the transformation of ligand 4  to ligand 7  are shown in Supplementary Figs.~S16-S19.

For these transformations, our results yielded an \text{RMSE} of $\text{0.42}$~kcal.mol\textsuperscript{-1} for ABFE and $\text{0.82}$~kcal.mol\textsuperscript{-1} for \text{RBFE}. These values represent an improvement over the benchmarks established by Steinbrecher et al.~\cite{steinbrecher2015accurate}, who reported an RMSE of $1.25\text{ kcal.mol\textsuperscript{-1}}$ for fragment-based systems. This excellent agreement further demonstrates the effectiveness of the \text{Dual-LAO} method for accurately modeling such fragment-like transformations.

\subsection{$\text{RBFEs}$ of CHK1 with Scaffold-Hopping Perturbations} \label{sec:CHK1}
As mentioned earlier, scaffold-hopping perturbations represent the most challenging and aggressive class of alchemical transformations studied using $\text{RBFE}$ methods. Scaffold hopping involves replacing the core structural motif (the 'scaffold') of a ligand with a structurally distinct motif while aiming to retain similar potency and the overall binding mode.

To further challenge our method, we utilized the CHK1 scaffold-hopping set from Ross et al.~\cite{ross2023maximal}. This set contains $\text{5}$ ligands involved in $\text{8}$ different transformations (the $\text{2D}$ structure of the ligands is available in Supplementary Fig.~S20 and the LoMap network illustrating the various alchemical connections are shown in Supplementary Fig.~S21). Fig.~\ref{fig:diff_systems} panels $\text{d}_{1}$ and $\text{d}_{2}$ present one such transformation, where the $\text{CH}_{2}\text{CH}_{3}$ moiety in the $\text{R}_{1}$ and $\text{R}_{2}$ positions of Ligand 19 transforms into a closed $\text{7}$-member ring in Ligand 21. This exemplifies a crucial ring opening/closing type of perturbation. The convergence of PMF and the $\lambda$ exploration over time for the transformation of ligand 19  to ligand 21  are shown in Supplementary Figs.~S22-S25.

For this difficult transformation class, the \text{Dual-LAO} method yielded an RMSE of $\text{0.31}$ kcal.mol\textsuperscript{-1} for \text{ABFE} and $\text{0.58}$ kcal.mol\textsuperscript{-1} for \text{RBFE}, indicating an excellent correlation between experimental and simulation results (see Fig.~\ref{fig:diff_systems}). These results show an improvement over the work of Ries et al.~\cite{ries2022relative}, who used Replica-Exchange Enveloping Distribution Sampling sampling procedure.

These results highlight the ability of the Dual-LAO approach to circumvent the numerical instabilities traditionally associated with ring-breaking transformations in single-topology frameworks. As noted by Liu et al. \cite{liu2015ring}, ring opening via dummy atoms often leads to convergence failures due to poorly defined bonded potentials at the alchemical endpoints. In contrast, our method utilizes a dual-topology scheme where the initial and final ligand states are defined as independent topologies. This architecture eliminates the need to break or close bond constraints within a single molecule, and Lambda-ABF-OPES mitigates the challenge of ensuring configurational overlap between independent topologies. This algorithm provides a robust biasing force that samples the joint alchemical-configurational space, ensuring that even during drastic structural transitions, the system remains numerically stable and the free energy converges rapidly through consistent sampling of the binding environment.

\begin{figure}[H]
	\centering
	\includegraphics[width=0.9\textwidth]{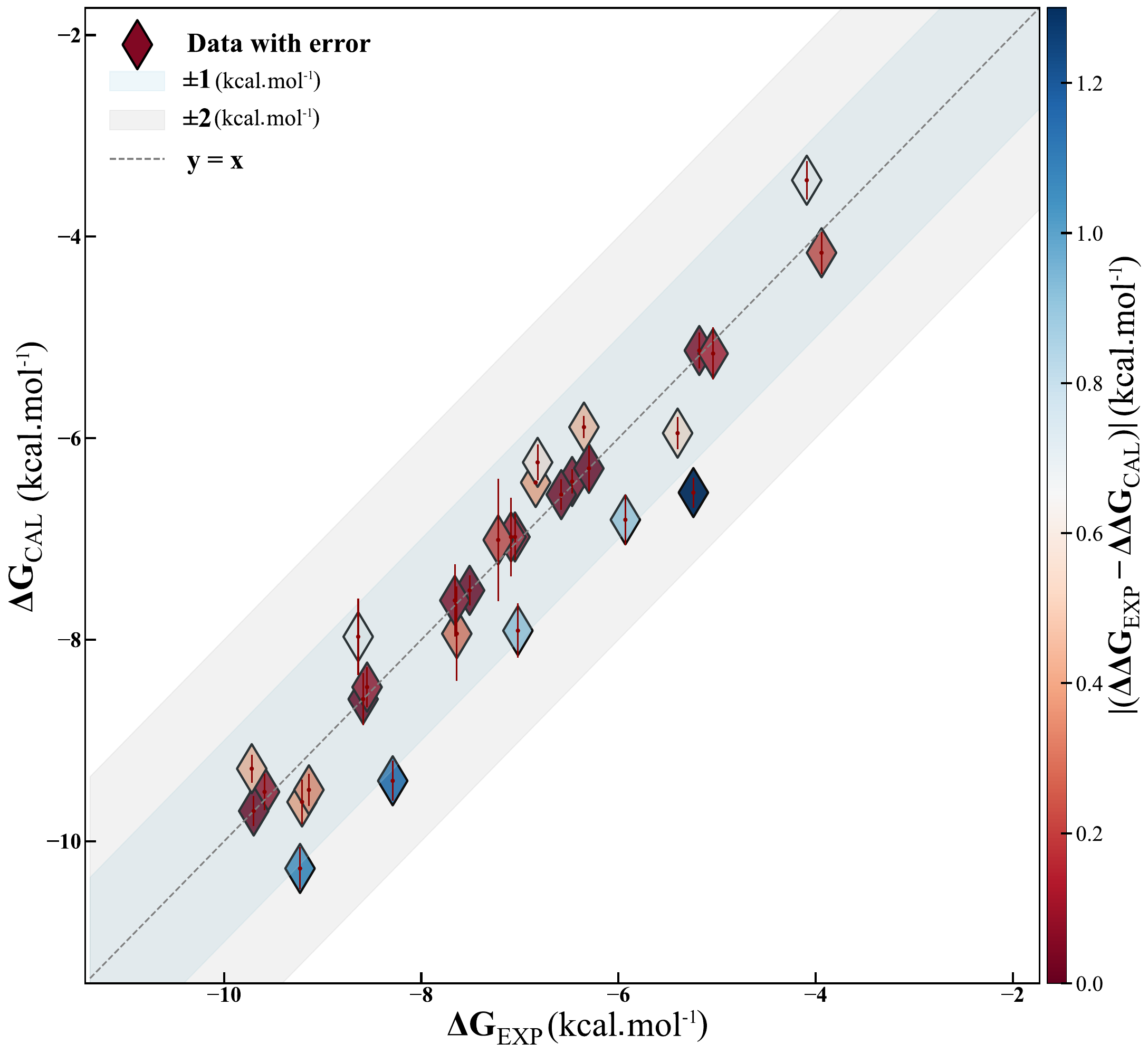}   
\caption{\textbf{Experimental vs. Calculated Absolute Binding Free Energies derived from the $\text{RBFE}$ network fit for all studied systems.} The calculated $\Delta G_{\text{CAL}}$ results and associated errors represent the mean $\pm$ standard error derived from the \text{Weighted Least-Squares ($\text{WLS}$) fitting} of the $\text{RBFE}$ network. The dark and light shaded regions indicate $\pm$1~kcal.mol\textsuperscript{-1} and $\pm$2~kcal.mol\textsuperscript{-1} deviations from the experimental values, respectively. The color bar displays the absolute difference between the experimental and computed values, $|\Delta G_{\text{EXP}} - \Delta G_{\text{CAL}}|$.}
	\label{fig:ABFE_All2}
\end{figure}

Table~\ref{tab:RBFE_Summary_Statistics} and Fig.~\ref{fig:ABFE_All2} collectively display the comprehensive performance of the $\text{Dual-LAO}$ method, covering the key statistics ($\text{R}^{2}$ and $\text{RMSE}$) across all four chemical systems, including both edgewise ($\Delta\Delta G_{\text{CAL}}$) and pairwise ($\Delta G_{\text{CAL}}$) $\text{RMSE}$ values.

\begin{table}[!ht]

    \centering

    \caption{\textbf{Summary of relative binding free energy ($\text{RBFE}$) results for all chemical systems.} \\
    Performance statistics of the \text{Dual-LAO} method compared to experimental data across all four chemical systems: PWWP1, BRD4, P38, and CHK1. The table includes the Protein Data Bank (\text{PDB}) identifier for the structure used, the total number of compounds and transformations (\text{edges}) in the set, the coefficient of determination ($\text{R}^{2}$) computed on $\Delta \Delta G$, and the \text{RMSE} calculated on both edgewise ($\Delta\Delta G_{\text{CAL}}$) and pairwise ($\Delta G_{\text{CAL}}$) free energy differences. All RMSE values are given in kcal.mol\textsuperscript{-1}.}

    \label{tab:RBFE_Summary_Statistics}

    \resizebox{\linewidth}{!}{%

        \begin{tabular}{ccccccc}

            \hline

            \textbf{System} & \textbf{PDB} & \textbf{No. compounds} & \textbf{No. edges} & $\mathbf{R^{2}}$ & \textbf{Edgewise RMSE} & \textbf{Pairwise RMSE} \\

            \hline

            \textbf{PWWP1} & 6G2B & 12 & 30 & 0.62 & 0.68 & 0.46 \\

            \hline

            \textbf{BRD4} & 5I88 & 8 & 11 & 0.76 & 0.42 & 0.64 \\ 

            \hline

            \textbf{P38} & 1W7H & 5 & 7 & 0.77 & 0.42 & 0.82 \\ 

            \hline

            \textbf{CHK1} & 3U9N & 5 & 8 & 0.48 & 0.31 & 0.58 \\ 

            \hline

            \textbf{Total} & - & 30 & 56 & 0.90 & 0.52 & 0.56 \\ \hline

            \hline

        \end{tabular}%

    }

\end{table}

\subsection{Multi-site Hydration Free Energies for Benzene Derivatives} 
We applied our multiple-topology approach to the computation of hydration free energies (HFE) of six molecules deriving from benzene, as a proof of concept. 
Similarly to the difference in binding affinity as calculated in Eq. (\ref{eq:ddgcycle}), here the difference in hydration free energy is computed as the difference between the free energies associated to the transformation of $L_1$ into $L_2$ in the solvent phase and in vacuum, such that:
\begin{equation}
\Delta\Delta G_{\text{hydration}} = \Delta G_{\text{solv}} - \Delta G_{\text{gas}}
\end{equation}

HFEs were computed on all fifteen possible pairs of molecules drawn from the chosen pool of six. 
Inputs were prepared by aligning molecules along the six carbons of the conjugated cycle. 
Multiple-topology transformations were explored over the course of 15~ns simulation.

As reference, HFEs were computed using the dual-topology method, with Dual-LAO. For each transformation, simulations of 2~ns were used for the solvent phase alchemical transformation, and 1~ns for the molecule in vacuum. For all transformations (dual and multiple topology), three replicas were run. 

A representation of the molecules (Supplementary Fig.~S26) as well as a comparison of the multi-site results with dual-site values (Supplementary Fig.~S27) are available in the Supplementary Information. Exploration of the expanded $\lambda$-space is shown in Supplementary  Fig.~S28, and demonstrates the ability of the method to navigate throughout the various edges of the alchemical transformations graph. 
The RMSE (dual vs. multiple topology) for all RBFE values was 0.22~kcal.mol\textsuperscript{-1}, and the Mean Average Error (MAE) 0.14~kcal.mol\textsuperscript{-1}, with an excellent correlation coefficient $R^2$ of 0.997. 
This preliminary result indicates the fidelity of the $\lambda$-jumping scheme that was used, and is intended to serve as a proof-of-concept for future methodological developments. 
In future work, we plan to apply this method to Binding Free Energies and will assess in particular its potential gain in efficiency (we expect exploration to be enhanced by the jumps) and in accuracy (the explicit exploration of cycles during such simulations should help more rigorous cycle closure when compared to sequences of independent simulations).

\subsection{Conclusion}

We presented \text{Dual-LAO}, a unified framework that addresses several long-standing bottlenecks in alchemical free-energy calculations by integrating \text{Lambda-ABF-OPES}, dual-topology, and dual-DBC restraints. The methodology's success and accelerated convergence across diverse benchmarks arise  from a synergistic mitigation of sampling and structural instabilities. By utilizing a dual-topology architecture, we circumvent the bonded-term complexities and numerical collapses typical of dummy-atom mapping, providing a smoother alchemical landscape for continuous $\lambda$ evolution. 

This structural stability is complemented by \text{dual-DBC} restraints, which prevent entropic sampling divergence in decoupled states, and the high-efficiency \text{ABF+OPES}$_e$ sampling engine. By adaptively biasing the configurational probability distribution, the framework effectively treats solvation dynamics and protein plasticity as orthogonal degrees of freedom. This allows for a concerted relaxation of the system that resolves the hysteresis traditionally associated with complex active-site rearrangements. Ultimately, \text{Dual-LAO} offers a numerically robust and rapidly converging approach, providing a practical path toward high-precision binding affinity predictions in challenging drug discovery targets.

Used in conjunction with the $\text{AMOEBA}$ polarizable force field, this strategy achieves chemical accuracy (overall $\text{RMSE}$ of 0.51 kcal.mol\textsuperscript{-1}) with a performance gain of more than an order of magnitude (over 15-fold acceleration) compared to state-of-the-art techniques. 
We demonstrated that the method is able to robustly tackle complex situations for which traditional approaches struggle. For example, we show that $\text{Dual-LAO}$ manages to reversibly rehydrate binding pockets upon ligand change (buried water displacement) without any specific tuning of the method. It also accurately handles other difficult perturbations including scaffold-hopping transformations, charge-changing, and ring-opening events with a marked improvement compared to state-of-the-art techniques.
These advances make $\text{Dual-LAO}$ a robust and highly efficient tool to drive hit-to-lead and lead-optimization in drug discovery.
The $\text{Multiple topology Dual-LAO}$ extension was successfully implemented and validated, confirming its potential for network-based $\text{RBFE}$ evaluations. Future work will focus on the systematic study of the sampling improvement obtained thanks to the multi-site extension (especially regarding cycle-closure), as well as the application of $\text{Dual-LAO}$ to machine-learning foundation models such as FeNNix-Bio1 \cite{ple2025foundation}.

\section{Methodology}

The RBFE is calculted following Eq. \ref{eq:ddgcycle} (see Ref. \citenum{pearlman_overlooked_1991} for the original thermodynamical cycle) which requires the computation of alchemical free energy components ($\Delta G_{\text{complex}}$ and $\Delta G_{\text{solv}}$). 
These components correspond to the transformation of a starting ligand ($L_1$) into a target ligand ($L_2$). 
Alchemical parameter $\lambda$ is used to scale key interactions in the system's Hamiltonian such that $\lambda=0$ corresponds to $L_1$ fully interacting and $\lambda=1$ corresponds to $L_2$ fully interacting (with $L_1$'s intermolecular interactions fully deactivated).

\subsection{Enhanced Alchemical Sampling: Dynamic $\lambda$ and $\text{Lambda-ABF-OPES}$}

In contrast to traditional discrete sampling techniques such as TI or FEP, utilizing fixed $\lambda$-points, this work treats $\lambda$ as a continuous dynamical variable within an extended Lagrangian framework. In this approach, $\lambda$ is allowed to explore the alchemical space $[0, 1]$ dynamically as a formal degree of freedom. This alchemical coordinate defines the transformation pathway through which electrostatic and van der Waals interactions are decoupled. Thus, while the evolution of $\lambda$ is continuous and dynamic, it follows a predefined functional mapping (interactions are scaled according to a specific schedule) to maintain numerical stability and ensure a singularity-free transformation across the entire range. The resulting free energy profile is subsequently reconstructed from the Potential of Mean Force (PMF) calculated along this continuous trajectory.

To ensure the required accuracy and rapid convergence, it is computed with the $\text{Lambda-ABF-OPES}$ framework \cite{ansari2025lambda}, which combines:

\begin{itemize}
    \item \textbf{Adaptive Biasing Force ($\text{ABF}$):} The $\text{Lambda-ABF}$ scheme applies an adaptive biasing force to the fictitious particle $\lambda$. This force enables proper sampling of the $\lambda$ space by flattening the free energy landscape along the alchemical path. The resulting $\text{PMF}$ is estimated by integrating the average force accumulated in 100 bins ($\Delta\lambda = 0.01$, see the Alchemical Sampling Parameters and Discretization section (Supplementary Methods) for further detail) using the $\text{TI}$ estimator.

    \item \textbf{$\text{OPES}$-Explore (OPES$_{e}$) :} The $\text{Lambda-ABF}$ scheme does not explicitly encompass capabilities to overcome energetic barriers orthogonal to the chosen collective variable ($\lambda$). This is addressed by the addition of the exploratory version of the On-the-fly Probability Enhanced Sampling (OPES)~\cite{Invernizzi2020,Invernizzi2022} method, known as OPES$_{e}$, a metadynamics-based method that aims at enhancing configurational sampling by adding a penalty to the already explored regions of the $\lambda$ interval (See the Alchemical Sampling Parameters and Discretization section (Supplementary Methods) for further detail).
\end{itemize}

To further accelerate convergence, \text{Lambda-ABF-OPES} is implemented using a multiple-walker parallelization scheme. In this approach, several independent replicas (walkers) of the system are simulated simultaneously. Crucially, these walkers share and contribute to a global bias potential and force-accumulation grid. This collective information exchange allows the algorithm to map the PMF more efficiently than a single-trajectory approach, as different walkers can explore distinct regions of the configurational space concurrently. 

In this study, we typically employed 4 walkers for the complex phase and 4 for the solvent phase. Consequently, a reported convergence time of $4$~ns/walker corresponds to a total aggregate sampling of $16$~ns for complex and $8$~ns for solvent. However, because these walkers run in parallel, the total wall-clock time remains equivalent to that of a $4$~ns MD simulation, representing a significant throughput advantage over serial alchemical methods.

This combination of efficient $\lambda$-space sampling from $\text{Lambda-ABF}$ and enhanced exploration across orthogonal barriers (such as complex solvent or water exploration~\cite{ansari2025lambda,blazhynska2025water}) from OPES$_{e}$ constitutes the $\text{Lambda-ABF-OPES}$ method.

\subsection{Dual-Topology Scheme and Restraints}

The transformation employs a \text{dual-topology scheme}~\cite{gao_hidden_1989,pearlman_overlooked_1991} where both ligands ($L_1$ and $L_2$) are simultaneously present. This approach simplifies input preparation (as no dummy atoms are required) and the computational cost of the extra atoms is negligible.

\begin{itemize}
    \item \textbf{$\text{MCS}$ Restraints:} The Maximal Common Substructure (MCS) is the substructure, common to both ligands, with the largest number of atoms or bonds.
    To ensure both ligands maintain a consistent conformation relative to their common parts, the MCS is identified using the FMCS algorithm\cite{dalke_fmcs_2013}, and a set of dual-topology restraints (harmonic wall restraints) is applied to corresponding atoms in the $\text{MCS}$.
    \item \textbf{Binding Pose Restraints:} To prevent the ligands from escaping the binding pocket while its intermolecular interactions are scaled down, a \text{dual-DBC scheme} was implemented to positionally restrain the ligands, as detailed below. 
\end{itemize}

\subsection{Alchemical Schedule} \label{sec:schedule}

\begin{figure}[h]
\centering
\includegraphics[width=0.7\textwidth]{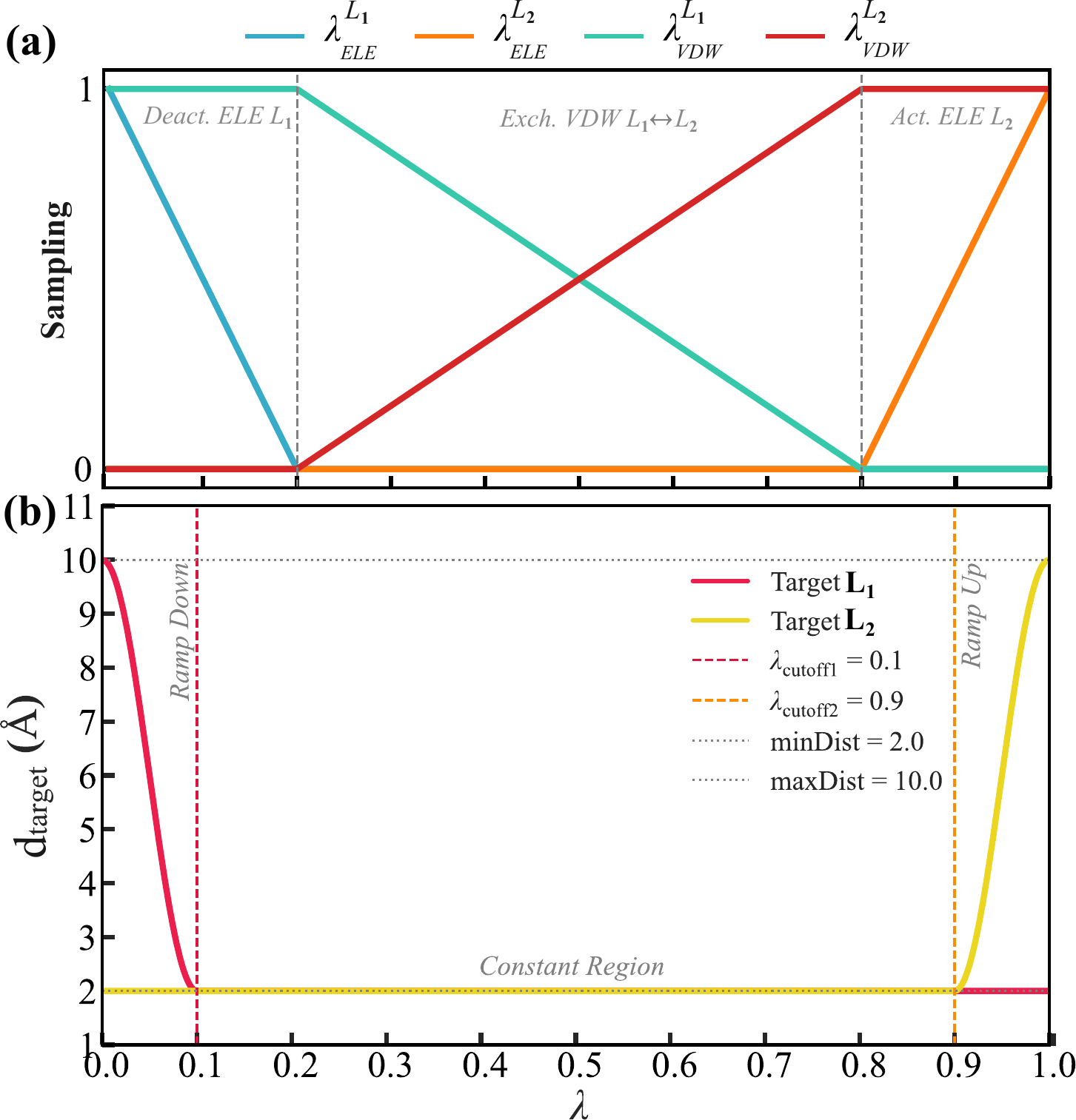}
\caption{ \textbf{Scaling of the intermolecular interactions and DBC restraint as a function of $\lambda$}. (a) Scaling of the intermolecular interactions for ligand $L_1$ ($\lambda_{\text{ELE}}^{L_1}$, $\lambda_{\text{VDW}}^{L_1}$) and ligand $L_2$ ($\lambda_{\text{ELE}}^{L_2}$, $\lambda_{\text{VDW}}^{L_2}$) as a function of $\lambda$. (b) Dual-DBC target profiles ($\mathbf{d_{\mathrm{target1}}}$ and $\mathbf{d_{\mathrm{target2}}}$).}
\label{fig:schedule}
\end{figure}

The alchemical transformation from $L_1$ to $L_2$ is controlled using scaling factors for the intermolecular interactions ($\lambda_{\text{ELE}}^X$ and $\lambda_{\text{VDW}}^X$). The continuous $\lambda$ schedule is divided into three steps, as illustrated in Fig. \ref{fig:schedule} panel (a):

\begin{enumerate}
    \item \textbf{Electrostatic Decoupling ($\lambda=[0, 0.2]$):} Electrostatic interactions of ligand $L_1$ are turned off by scaling down linearly its permanent multipoles and polarizability.
    \item \textbf{VDW Switch ($\lambda=[0.2, 0.8]$):} Van-der-Waals interactions for $L_1$ are linearly turned off, and the same interactions for $L_2$ are linearly turned on.
    \item \textbf{Electrostatic Coupling ($\lambda=[0.8, 1.0]$):} Electrostatic interactions of ligand $L_2$ are linearly activated by scaling up linearly its permanent multipoles and polarizabilities.
\end{enumerate}
It can be noted that over the "VDW Switch" section, all electrostatic interactions of both ligands are turned off. If ligands bound pose is mainly stabilized through electrostatics, positional drift can then be observed. This caveat is addressed through the use of DBC restraints, as was the case in the original Lambda-ABF-OPES approach\cite{ansari2025lambda}, described in the following paragraphs.

\subsection{Dual DBC Restraint Scheme } \label{sec:DualDBC}

To counteract positional drift during the intermediate $\lambda$ states, we apply a flat-bottom harmonic restraint on the $\text{DBC}$ of each ligand, using two independent, $\lambda$-dependent targets, $d_{\mathrm{target1}}$ and $d_{\mathrm{target2}}$, as shown in Fig.~\ref{fig:schedule}(b).

The restraint acts only if $\text{DBC}$ is higher than the target. The target are designed to have a minimum $D_{\text{min}} = \SI{2.0}\,{\angstrom}$ and a maximum $D_{\text{max}} = \SI{10.0}\,{\angstrom}$.

\begin{align}
d_{\mathrm{target1}}(\lambda) &=
\begin{cases}
D_{\text{min}} & \text{for } 0.0 \le \lambda \le 0.9 \\
D_{\text{min}} + \left( \frac{D_{\text{max}} - D_{\text{min}}}{2} \right) \left[ 1 - \cos\left( \frac{\pi (\lambda - 0.9)}{0.1} \right) \right] & \text{for } 0.9 < \lambda \le 1.0
\end{cases}
\label{eq:dtarget1} \\
d_{\mathrm{target2}}(\lambda) &=
\begin{cases}
D_{\text{min}} + \left( \frac{D_{\text{max}} - D_{\text{min}}}{2} \right) \left[ 1 + \cos\left( \frac{\pi \lambda}{0.1} \right) \right] & \text{for }0 \le \lambda < 0.1 \\
D_{\text{min}} & \text{for } 0.1 \le \lambda \le 1.0
\end{cases}
\label{eq:dtarget2}
\end{align}

Restraints are maintained at $D_{\text{min}}$, but are gradually released near the end point corresponding to the decoupled ligand of interest by increasing the target to $D_{\text{max}}$. As explained in a further section, the value of $D_{\text{min}}$ is chosen so that the binding mode of the fully coupled ligand of interest is not biased. This ensures that the final (fully coupled and fully decoupled) states are unbiased (see Drivers of Protocol Efficiency and Stability section (Supplementary Methods) for further detail).

\subsection{Implementation}

The $\text{Dual-LAO}$ methodology was implemented within the \text{Tinker-HP} molecular dynamics software package~\cite{adjoua_tinker-hp_2021}, leveraging the capabilities of the \text{Colvars} library \cite{fiorin2024expanded} for managing the enhanced sampling Collective Variables and the specialized restraints.

\subsection{Simulation Details}
\subsubsection{Plain MD Simulations}
In this work, all molecular dynamics (MD) and advanced sampling simulations were performed using the Tinker-HP molecular dynamics package (Version 1.2)~\cite{tinker-hp,adjoua_tinker-hp_2021}, which includes the integrated \texttt{Colvars}~\cite{fiorin2013using} library for $\text{Dual-LAO}$ $\text{RBFE}$ calculations. 
Simulations utilized the polarizable AMOEBA force field 
\cite{ponder_current_2010,wu2012automation,shi_polarizable_2013,AMOEBA-nucleic} for the protein, solvent (water), and counter-ions. Ligand parameters were derived using the \texttt{Poltype} package \cite{poltype2} (see Computational Cost and Performance Metrics section (Supplementary Methods) for further detail). Each system was constructed within a cubic water box, sized to ensure a minimum separation of $12$~\AA{} between the ligand and any box boundary. The solvent was set to an aqueous $\text{NaCl}$ solution at a concentration of $0.15$~M to ensure physiological relevance and system neutrality. 

A multi-stage protocol was employed to prepare the systems for production dynamics. First, the system underwent energy minimization via the Tinker-HP $\mathtt{minimize}$ utility, followed by a two-step heating phase from $200$~K to $298$~K under the NVT (canonical) ensemble. The first two steps utilized the RESPA integrator with a $2.0$~fs time step: Step~1 ($4$~ns) involved applying positional restraints to all protein, ligand, and structural X-ray water atoms; in Step~2 ($4$~ns), these restraints were relaxed to act only on the heavy backbone atoms of the protein, maintained with a force constant of $10$~$\text{kcal.mol\textsuperscript{-1}.\AA\textsuperscript{-2}}$. Next, the system underwent a two-step relaxation phase. Step~3 ($5$~ns) involved further equilibration in the NVT ensemble using the three-level BAOAB-RESPA1 multiple-timestep integrator \cite{lagardere_pushing_2019}, where the outer time step was $10$~fs, with intermediate and inner steps of $10/3$~fs and $1$~fs, respectively. For this step, restraints were applied only to $\text{C}\alpha$ atoms using a reduced force constant of $1$~$\text{kcal.mol\textsuperscript{-1}.\AA\textsuperscript{-2}}$. Finally, Step 4 ($5$~ns) conducted the final equilibration under the NPT (isothermal-isobaric) ensemble at $1$ atm pressure, maintaining the $\text{C}\alpha$ restraints from the previous step. Following equilibration, a $20$ ns production run was executed for each ligand complex in the NPT ensemble. The $\text{BAOAB-RESPA1}$ Langevin integrator was used with a $10$ fs outer time step. Temperature control in non-Langevin steps was ensured by the $\text{Bussi}$ thermostat \cite{bussi2007canonical}, while pressure was regulated by the $\text{Berendsen}$ barostat \cite{berendsen1984molecular}. Van der Waals interactions employed a $12$~\AA{} cutoff. Electrostatic interactions were computed using the Particle Mesh Ewald (PME) method \cite{essmann1995smooth} with a $7$~\AA{} real-space cutoff. Induced dipoles were determined via a Preconditioned Conjugate Gradient (PCG) solver \cite{lagardere2015scalable} to a convergence tolerance of $1 \times 10^{-5}$~Debye.

The computational cost associated with force field parameterization and production simulations is detailed in the \textit{Computational Cost and Performance Metrics} section of the Supporting Information. For representative systems ($\approx$60K atoms), production throughput reached up to 55~ns per day depending on the hardware architecture (see Table S4 in the SI).

\subsubsection{$\text{Dual-LAO}$ Simulations}

For all alchemical simulations, including R-group, buried water displacement, scaffold-hopping, and fragment-like transformations, we employed the $\text{Dual-LAO}$ method. We ran four walkers for each transformation, with 4~ns per walker for the complex phase and 2~ns per walker for the solvent phase. All other simulation parameters matched those of the plain MD simulations, with one exception: we used the BAOAB-RESPA integrator~\cite{lagardere_pushing_2019} with a 2 fs time step.

\subsection{DBC Selection}

As described in Dual DBC Restraint Scheme section, we employed the $\text{DBC}$~\cite{henin2018dbc} to restrain each $L_1$ and $L_2$ within its respective binding pose during complex alchemical transformations. The DBC coordinate quantifies the ligand's deviation from its target pose by calculating the RMSD of a selected set of ligand atoms (noted LA) after optimal superposition onto a reference set of receptor binding site atoms (noted RB). 
This single variable efficiently captures positional, orientational, and conformational fluctuations of the ligand relative to its binding environment. 

The selection of $\text{LA}$ and $\text{RB}$ atoms is crucial for defining a meaningful DBC restraint. To identify the least mobile parts of the ligand, we monitored the Root Mean Square Fluctuation ($\text{RMSF}$) of all heavy ligand atoms over a $20$~ns conventional MD simulation of each individual complex. For the selection of the ligand atoms ($\text{LA}$) used in the $\text{DBC}$ restraint, we applied a uniform Root Mean Square Fluctuation ($\text{RMSF}$) cutoff of 1.0~\AA \ across all ligands and targets under study. This ensures consistency in restraining the least flexible portions of the ligands undergoing alchemical coupling/decoupling. The $\text{RB}$ set consisted of the $\text{C}\alpha$ atoms of the protein that fell within $6$~\AA{} of the ligand and exhibited an intrinsic flexibility below approximately $0.8$~\AA{} during the production MD.

The use of these uniform RMSF cutoffs serves to standardize the protocol and minimize operator bias across diverse targets. By selecting atoms based on a fixed flexibility threshold rather than manual inspection, the process becomes largely automated and reproducible. The sensitivity of the resulting free energies to the exact composition of the LA and RB sets is minimized by the nature of the DBC variable itself; as a collective RMSD-based restraint, it is inherently more stable than restraints based on individual atom pairs. Consequently, provided that the selection captures the relatively rigid core of the binding interface, minor variations in the atom sets do not significantly alter the sampled phase-space volume.

To ensure that the DBC restraint does not introduce artificial bias during the alchemical simulation, we implemented a dual-DBC restraint strategy as explained earlier. Specifically, a minimum DBC value ($\text{DBC}_{\text{min}}$) was set at $2$~\AA{}. This $\text{DBC}_{\text{min}}$ value was chosen to be higher than the upper bound of DBC coordinate value observed over the plain MD production run for each ligand, which was monitored using the $\text{Colvars}$ library. A flat-bottomed harmonic restraint was applied to the $\text{DBC}$ coordinate of both $L_1$ and $L_2$ during the $\text{Dual-LAO}$ simulations, following the DBC restraint schedule. This restraint employed a high force constant of $100$ $\text{kcal.mol\textsuperscript{-1}.\AA\textsuperscript{-2}}$ and was active only when the calculated $\text{DBC}$ value exceeded the predetermined $\text{DBC}$ cutoff.

\subsection{Multiple Topology $\text{RBFEs}$}

The dual-topology framework is readily extensible to a multiple-topology framework (akin to the Separated Topologies Approach \cite{baumann2023broadening}), which allows several alchemical transformations to be explored over the course of a single simulation. In a typical drug discovery campaign, the goal is to explore a network of transformations connecting a series of ligands ($L_1, L_2, L_3, \dots$). This network is defined by the user (e.g., using methods like $\text{LoMap}$ \cite{liu2013lead}) and is naturally represented as a graph. In this representation, each node denotes a molecule, and each edge represents the alchemical transformation between two connected molecules (not all pairs of molecules must be connected).
Crucially, while the dual-topology framework explores each edge of this graph during an independent simulation, the multiple-topology framework allows all connected transformations to be explored during a single, extended simulation, following the procedure detailed below.

Any alchemical state of the multiple-topology simulation can be characterized with the triplet ($L_i$, $L_j$, $\lambda$), where $L_i$ and $L_j$ are the two molecules undergoing alchemical transformation, and $\lambda$, comprised in $[0;1]$, describes the evolution of the transformation (Following this notation, the dual-topology case is simply described as the triplet $(L_1,L_2,\lambda)$). If $\lambda=0$, then $L_i$ is fully coupled (and $L_j$ fully decoupled), and if $\lambda=1$, $L_i$ is decoupled, $L_j$ coupled.

The $\text{Dual-LAO}$ simulation is started on an (arbitrarily chosen) initial edge $(L_i,L_j)$ of the graph. Just like in the dual-topology case, the value of $\lambda$ is used to explore the alchemical schedule and define scaling terms for the VDW and ELE interactions of $L_i$ and $L_j$, while all other ligands remain fully decoupled. 

In the dual-topology scheme, the value of $\lambda$ is kept within the $[0;1]$ segment using reflexive boundary conditions.
In the multiple-topology scheme, after applying reflexive boundary conditions, a special procedure is applied to enable exploration of other edges:
\begin{enumerate}[label=(\roman*)]
\item The node corresponding to the reflection value is identified.
\item  All edges connected to this node are identified. 
\item  A random integer between 1 and the number of connected edges is drawn. This chooses the next edge of the network to explore, and defines a new triplet-state to explore.
\end{enumerate}

This process allows the simulation to explore a new alchemical state whenever $\lambda$ reaches (or exceeds) 0 or 1, which corresponds a ``fully coupled''  alchemical state.

Let us illustrate the process with an example: the simulation starts at $(L_1, L_2, \lambda=0)$. After some time, $\lambda$ reaches $1$, which corresponds to the ligand $L_2$ being fully coupled. Let's assume $L_1, L_3, L_4$ are connected to $L_2$ in our graph. A random integer is drawn between 1 and 3 (say, 2). This chooses the second connected edge to continue exploration, here $L_2$ to $L_3$. The new triplet-state becomes $(L_3, L_2, \lambda=1)$, and now alchemical transformation from $L_2$ to $L_3$ is explored. 

With minimal extra input from the user (namely, the choice of a network via a connectivity matrix), this method allows to compute several $\text{RBFEs}$ over the course of a single simulation.

\section{Data Availability Statement}
The Tinker-HP and colvars input files used, and the structures
are available on GitHub at:  
\href{https://github.com/ansarinarjes/RBFE-L-ABF-OPES.git}{https://github.com/ansarinarjes/RBFE-L-ABF-OPES.git}

\section{Code availability}
The code used during the study is available on GitHub: 

\href{https://github.com/TinkerTools/tinker-hp}{https://https://github.com/TinkerTools/tinker-hp}

\href{https://github.com/TinkerTools/poltype2}{https://https://github.com/TinkerTools/poltype2}

\href{https://github.com/Colvars/colvars}{https://github.com/Colvars/colvars}

\section*{Conflict of interest/Competing interests} 
L. L., and J.-P. P. are co-founders and shareholders of Qubit Pharmaceuticals. The remaining authors declare no competing interests.

\section{Contributions}
\textbf{Designed research:} L. L.; 
\textbf{Performed research:} N. A., F. A., L. L. ;
\textbf{Performed simulations:} N. A., F. A. ;
\textbf{Contributed analytic tools:}  N. A., F. A.; 
\textbf{Contributed new code:}  F. A., L. L.; 
\textbf{Analyzed data:} N. A., F. A., J. H.,  L. L., J.-P. P.; 
\textbf{Wrote the paper:} N. A., F. A., J. H., L. L., J.-P. P.

\section{Acknowledgment}
The authors would like to extend their gratitude to Oliver Adjoua and Florent Hédin for their unwavering support in all matters of code development, as well as Chengwen Liu for his help with AMOEBA parametrization. 

This work was supported by computing grants from the Grand Équipement de Calcul Intensif ($\text{GENCI}$), Institut du Développement et des Ressources en Informatique ($\text{IDRIS}$), and Centre Informatique de l’Enseignement Supérieur ($\text{CINES}$), France, under grant numbers AD010715770 and AD010716167R1.

\bibliography{General}

\end{document}


\clearpage
\begin{center}
\textsuperscript{$\nabla$} N.A. and F.A. contributed equally.   
\end{center}

\clearpage

\subsection{PWWP1}

\begin{figure}[!h]
	\centering
	\includegraphics[width=0.8\textwidth]{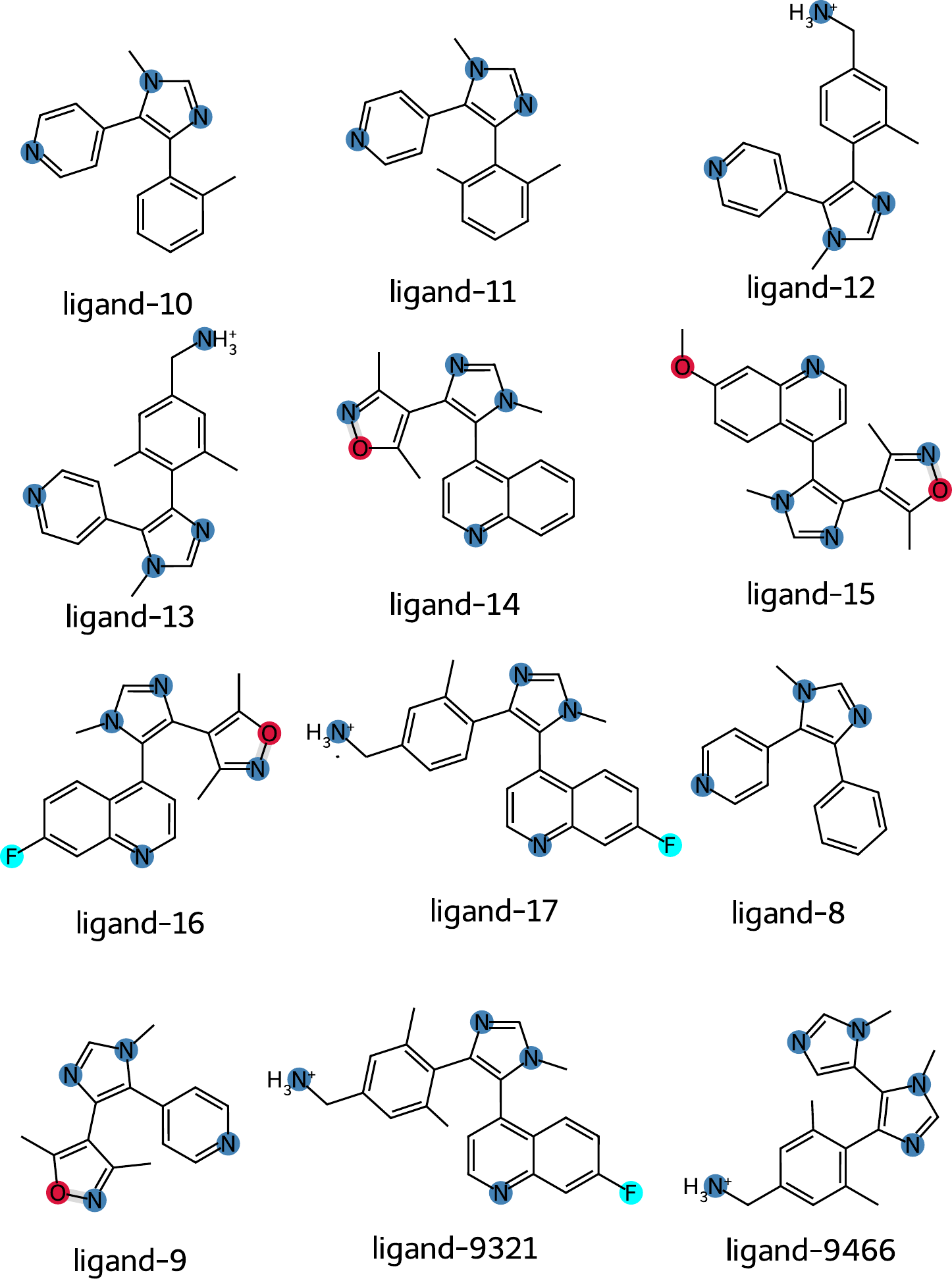}   
\caption{\textbf{2D structures of all ligands complexed with PWWP1.}}
	\label{fig:lig_PWWP1}
\end{figure}

\begin{figure}[!h]
	\centering
	\includegraphics[width=0.9\textwidth]{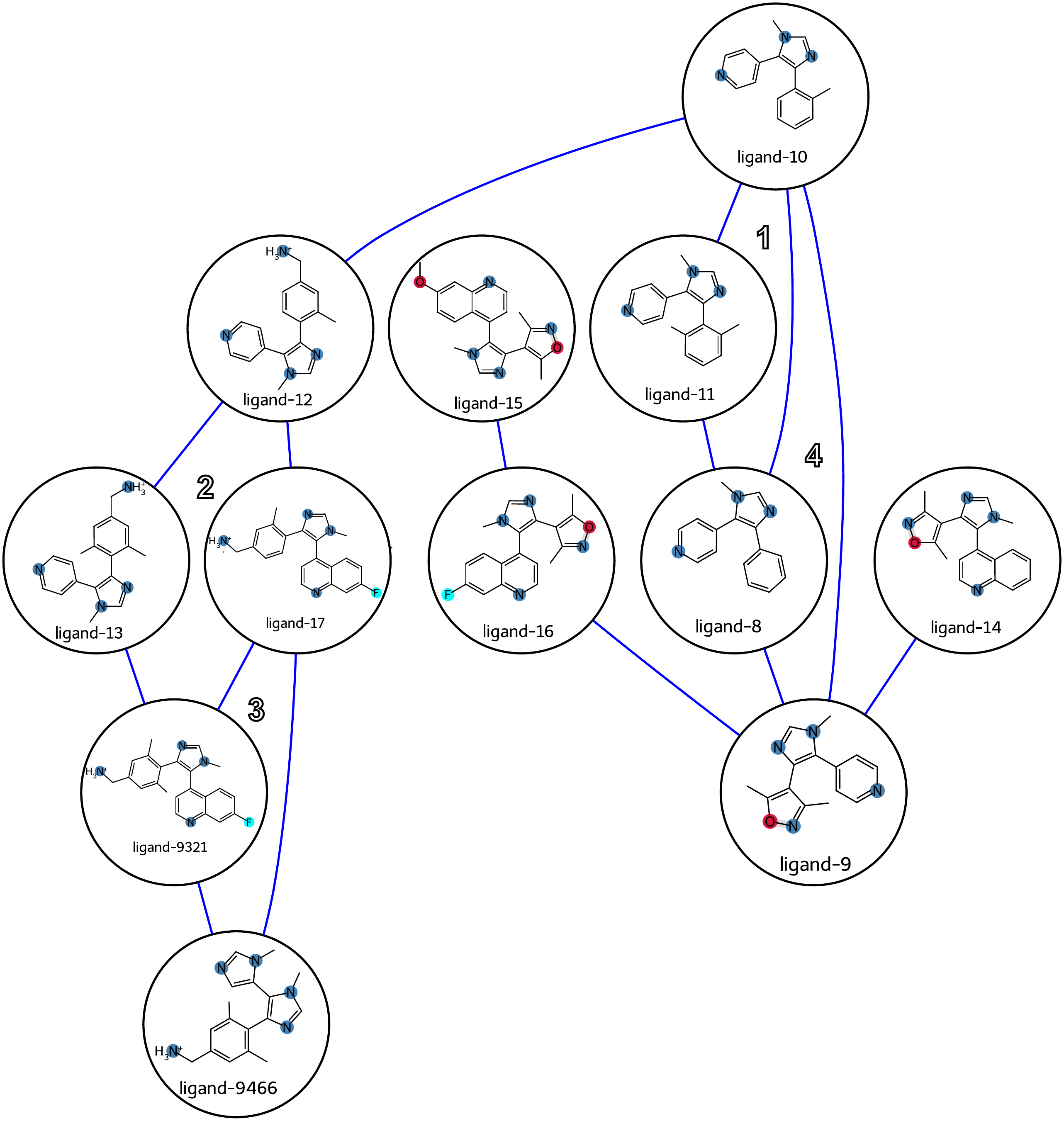}   
\caption{\textbf{LoMap perturbation network for PWWP1 ligand pairs}. Nodes represent individual ligands, and edges indicate the simulated transformation paths. Numeric labels centered within each loop identify the different three-member closed cycles.}
	\label{fig:lomap_PWWP1}
\end{figure}

\begin{figure}[!h]
	\centering
	\includegraphics[width=0.6\textwidth]{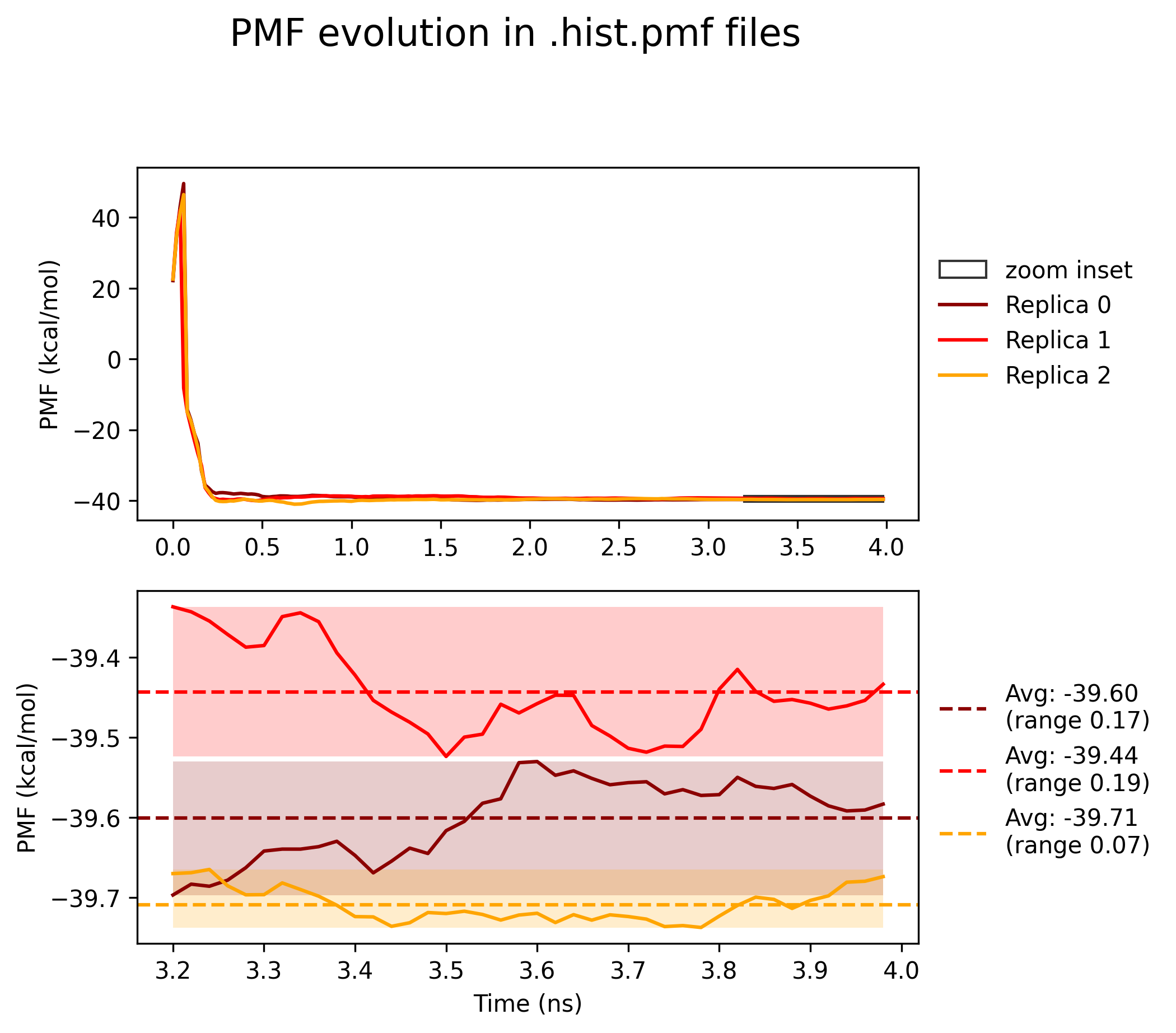}   
\caption{\textbf{Convergence plot of the Potential of Mean Force (PMF) for the complex phase of the ligand 10 to ligand 12 transformation in PWWP1.}}
	\label{fig:pmf_com_charge_change}
\end{figure}

\begin{figure}[!h]
	\centering
	\includegraphics[width=0.9\textwidth]{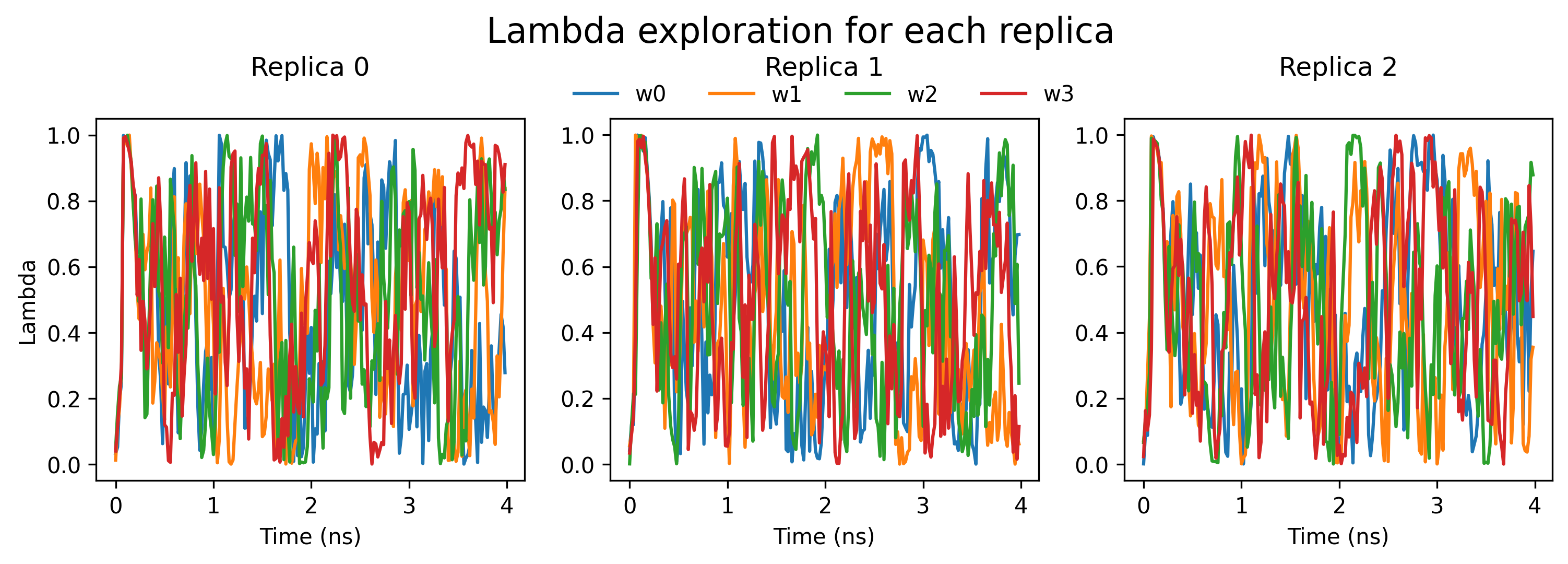}   
\caption{\textbf{Lambda fluctuation over time for the complex phase of the ligand 10 to ligand 12 transformation in PWWP1.}}
	\label{fig:lam_com_charge_change}
\end{figure}

\begin{figure}[!h]
	\centering
	\includegraphics[width=0.6\textwidth]{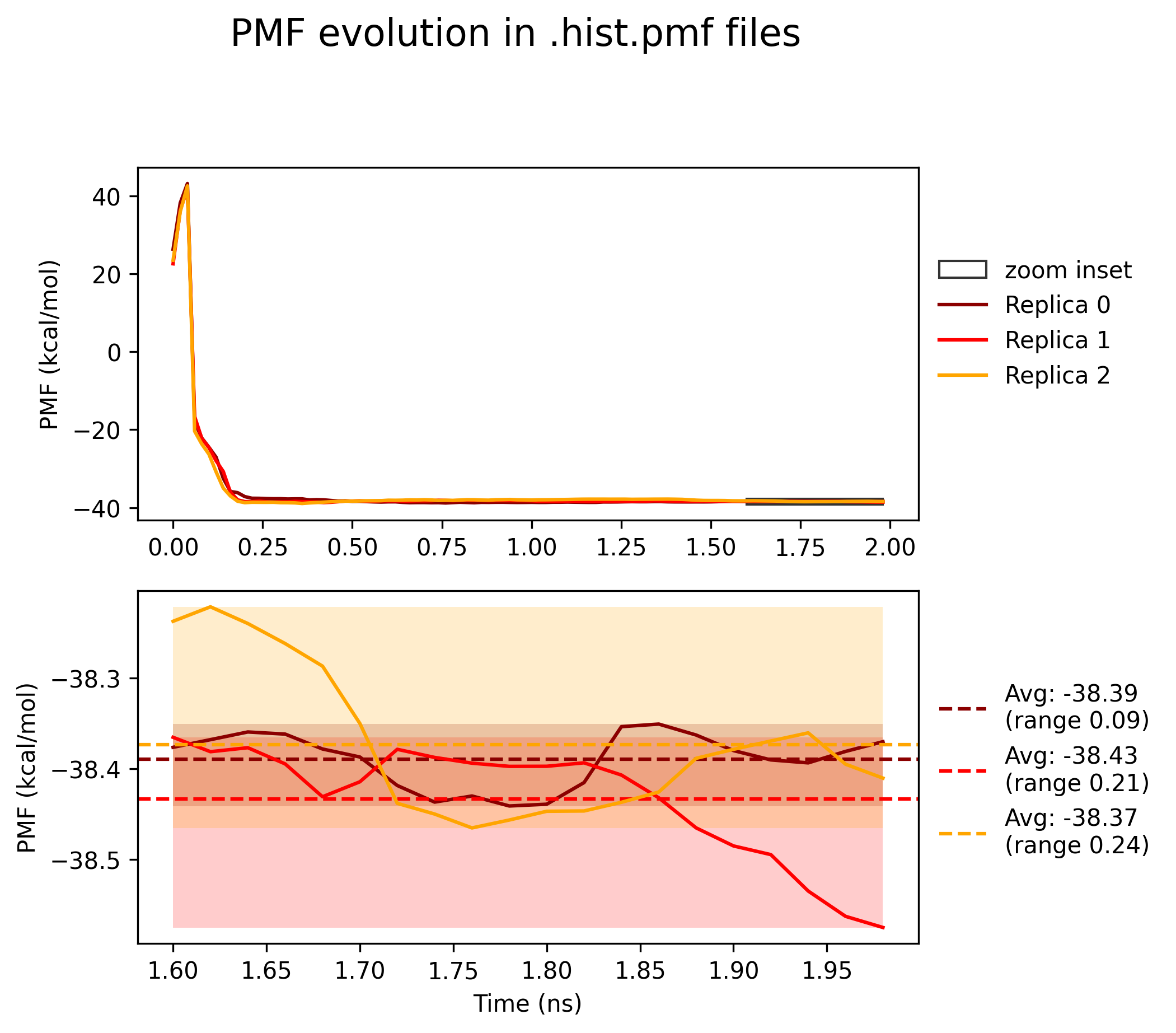}   
\caption{\textbf{Convergence plot of the Potential of Mean Force (PMF) for the solvent phase of the ligand 10 to ligand 12 transformation in PWWP1.}}
	\label{fig:pmf_sol_charge_change}
\end{figure}

\begin{figure}[!h]
	\centering
	\includegraphics[width=0.9\textwidth]{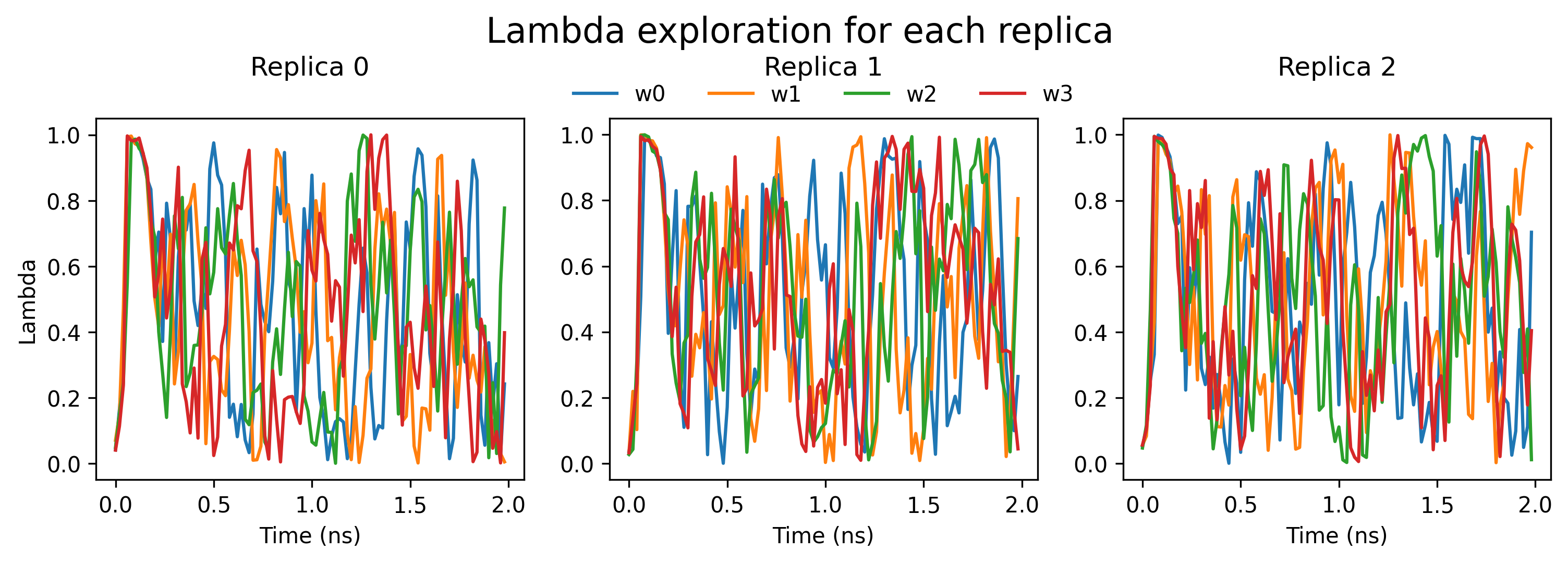}   
\caption{\textbf{Lambda fluctuation over time for the solvent phase of the ligand 10 to ligand 12 transformation in PWWP1.}}
	\label{fig:lam_sol_charge_change}
\end{figure}

~

\begin{table}[!ht]
    \centering
    \begin{tabular}{l m{7em} m{3em} c}
    \hline
        \textbf{Ligand IDs} & \textbf{Cycle closure (kcal.mol\textsuperscript{-1})} & \textbf{Cycle length} & \textbf{Label (Fig. \ref{fig:lomap_PWWP1})} \\ \hline
        \textbf{11 - 8 - 10} & 0.32 $\pm$ 0.52 & 3 & \textbf{(1)} \\ \hline
        \textbf{17 - 12 - 13 - 9321} & 0.23  $\pm$ 0.89 & 4 & \textbf{(2)} \\ \hline
        \textbf{17 - 9321 - 9466} & -0.61  $\pm$ 0.95 & 3 & \textbf{(3)} \\ \hline
        \textbf{8 - 9 - 10} & 0.29  $\pm$ 0.85 & 3 & \textbf{(4)} \\ \hline
    \end{tabular}
    \caption{\textbf{Cycle closures values for PWWP1} computed on 3- and 4-edges long cycles. First column reports ligand numbers constituting the cycle. Correspondence with the cycles in the LoMap figure given through label in last column. }
    \label{tab:cycle_closure}
\end{table}

\clearpage

\section{Assessment of Charge-Change Corrections}

To evaluate the necessity of explicit charge-change correction schemes (e.g., as described by Chen et al.~\cite{wu2022correction}), we performed a validation study on the neutral-to-charged transformation of lig10 $\to$ lig12 in PWWP1 ($q = 0 \to +1$). 

We employed a co-alchemical ion approach to maintain net neutrality of the simulation cell throughout the transformation. A chloride (Cl$^{-}$) dummy ion was introduced and coupled to the alchemical state of the ligand; as the positive charge of lig12 was decoupled ($\lambda \to 0$), the Cl$^{-}$ ion was simultaneously decoupled. This ensures that the net charge of the periodic box remains zero at both alchemical endpoints.

\begin{table}[h]
\centering
\caption{Comparison of $\Delta\Delta G$ values (kcal.mol\textsuperscript{-1}) for the lig10 $\to$ lig12 transformation with and without charge-neutralization corrections.}
\begin{tabular}{l c}
\hline
Method & $\Delta\Delta G$ (kcal.mol\textsuperscript{-1}) \\
\hline
Experimental & $-1.04$ \\
Dual-LAO (Standard) & $-1.22 \pm 0.11$ \\
Dual-LAO (Co-alchemical Ion) & $-0.90 \pm 0.31$ \\
\hline
\end{tabular}
\label{tab:charge_correction}
\end{table}

As shown in Table~\ref{tab:charge_correction}, the results obtained using the co-alchemical ion scheme are in high agreement with both the experimental value and our original results. The overlapping error bars and the small magnitude of the shift ($< 0.3$ kcal.mol\textsuperscript{-1}) suggest that for the PWWP1 system, the combination of a large simulation box and the enhanced sampling of the Dual-LAO framework effectively minimizes the artifacts typically associated with charge-changing perturbations.

\begin{figure}[!h]
	\centering
	\includegraphics[width=0.9\textwidth]{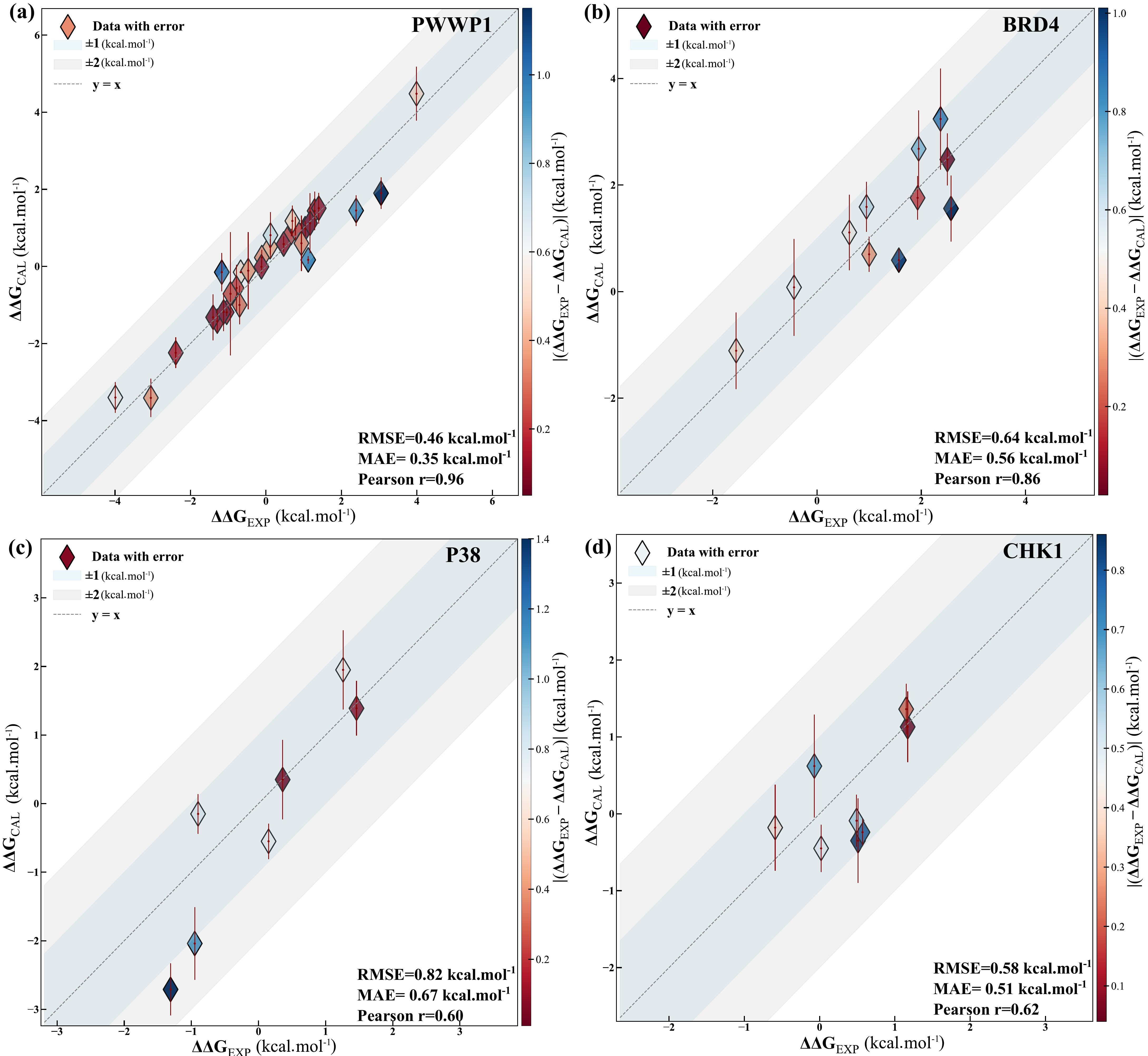}   
\caption{\textbf{Experimental vs. Calculated Relative Binding Free Energy Changes.} (a) PWWP1, (b) BRD4, (c) P38, and (d) CHK1 complexes. The calculated $\Delta\Delta G_{\text{CAL}}$ values from RBFE results and associated errors represent the mean $\pm$ standard error from three independent repeats. The dark and light shaded regions indicate $\pm$1~kcal.mol\textsuperscript{-1} and $\pm$2~kcal.mol\textsuperscript{-1} deviations, respectively. The color bar displays the absolute difference between the experimental and computed values, $|\Delta\Delta G_{\text{EXP}} - \Delta\Delta G_{\text{CAL}}|$. For each system, Pearson's $r$, RMSE, and MAE statistics are reported.}
	\label{fig:RBFE_All}
\end{figure}

\clearpage

\subsection{BRD4}

\begin{figure}[!h]
	\centering
	\includegraphics[width=0.9\textwidth]{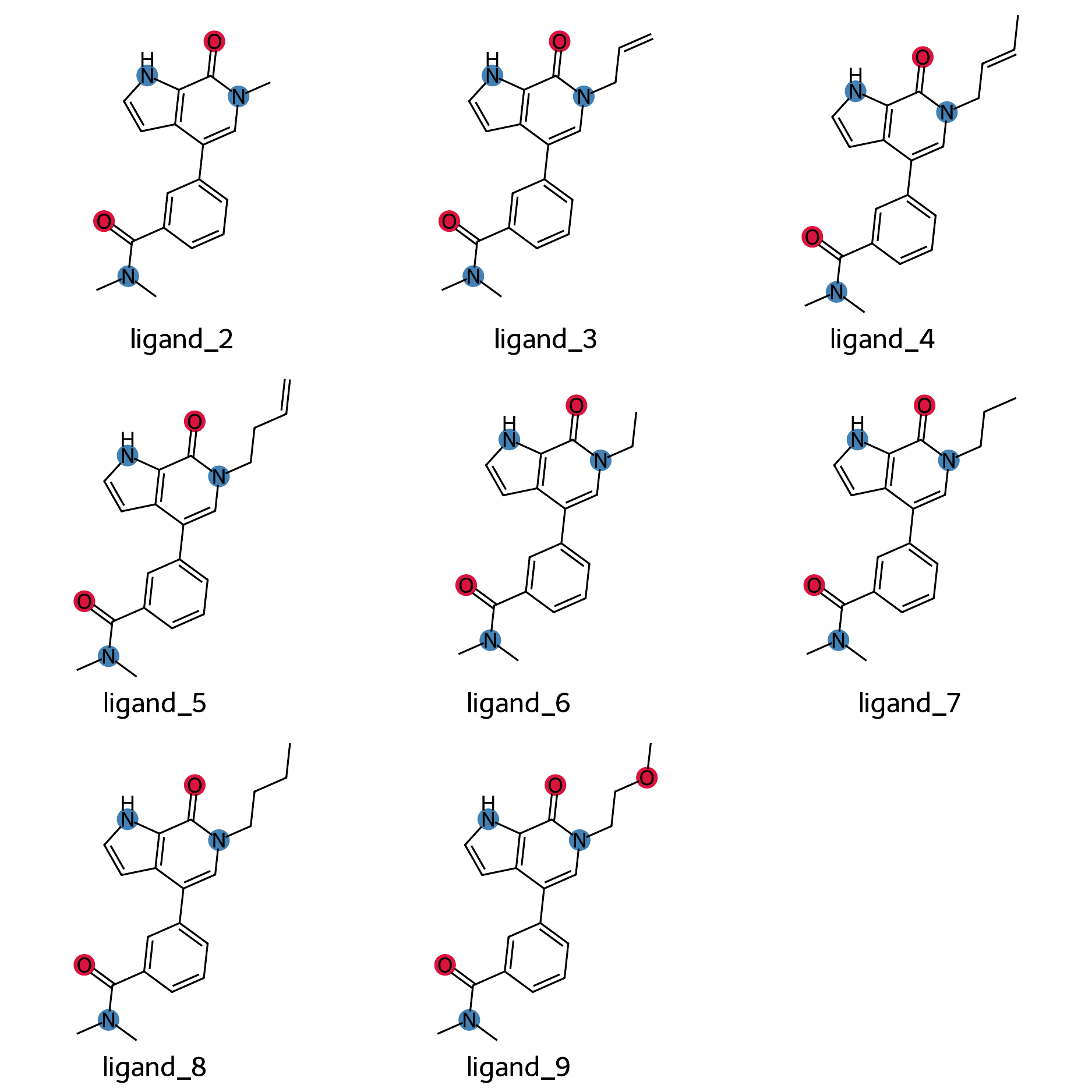}   
\caption{\textbf{2D structures of all ligands complexed with BRD4.}}
	\label{fig:lig_BRD4}
\end{figure}

\begin{figure}[!h]
	\centering
	\includegraphics[width=1.0\textwidth]{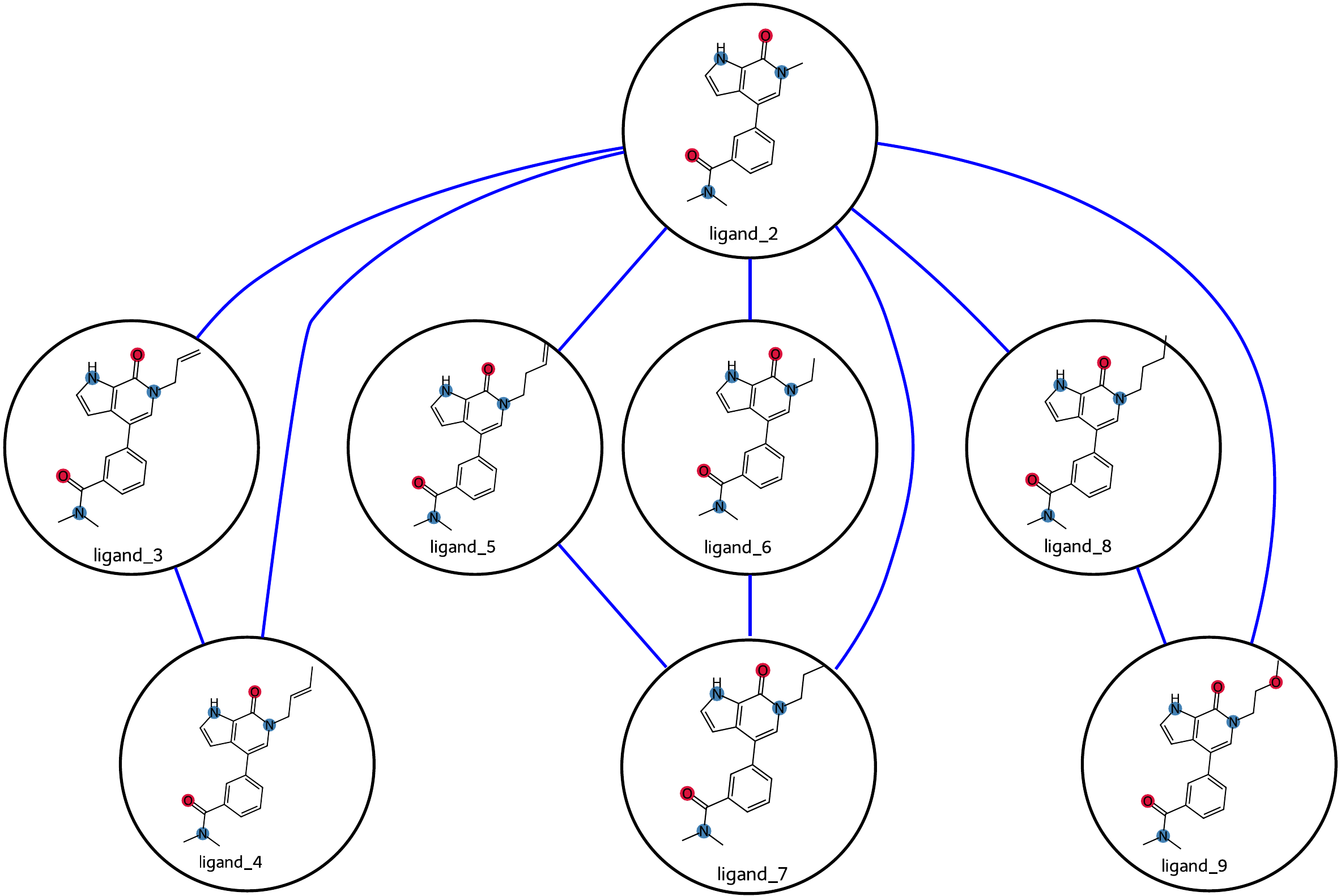}   
\caption{\textbf{LoMap perturbation network for BRD4 ligand pairs}. Nodes represent individual ligands, and edges indicate the simulated transformation paths.}
	\label{fig:lig_brd4_lomap}
\end{figure}


\begin{figure}[!h]
	\centering
	\includegraphics[width=0.9\textwidth]{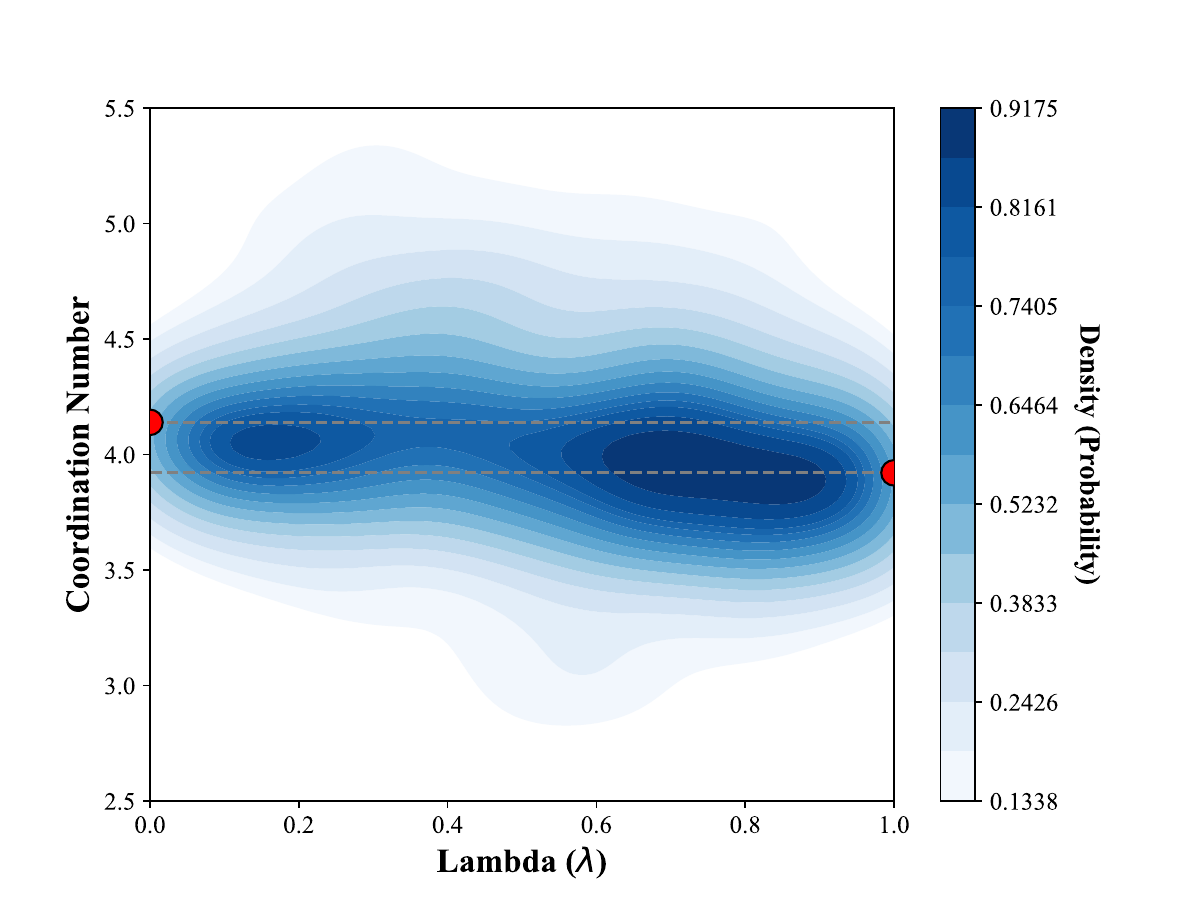}   
	\caption{\textbf{Density Plot of Water Coordination Number around Ligand $\text{O1}$ Atom During Alchemical Transformation}. The plot shows the density distribution of the coordination number ($\text{CN}$) of water molecules around the $\text{O1}$ atom of $\text{ligand 2}$ as a function of the alchemical coupling parameter, $\lambda$, for the $\text{ligand 2} \to \text{ligand 5}$ transformation. Water oxygen atoms ($\text{O}_{\text{water}}$) were counted within a cutoff distance of $5 \text{ \AA}$ from the $\text{O1}$ atom of the transforming ligand. The red circles mark the average coordination numbers at the end states: $\lambda=0$ (representing $\text{ligand 2}$) and $\lambda=1$ (representing $\text{ligand 5}$). The corresponding dashed lines indicate these average coordination number values, which slightly decrease as the system transforms from $\text{ligand 2}$ to $\text{ligand 5}$.}
    \label{fig:coordination_number_density} 
\end{figure}

\begin{figure}[!h]
	\centering
	\includegraphics[width=0.6\textwidth]{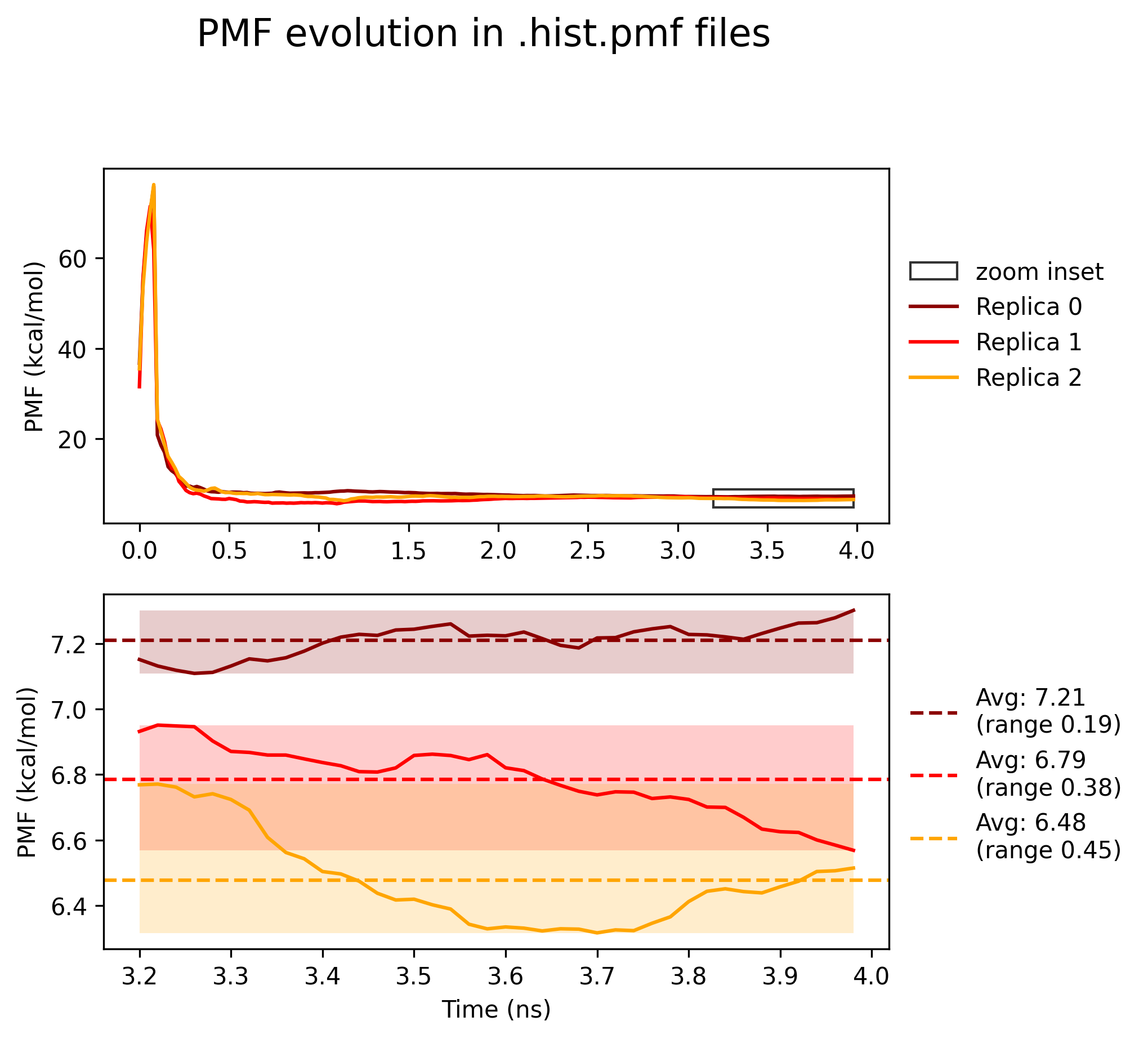}   
\caption{\textbf{Convergence plot of the Potential of Mean Force (PMF) for the complex phase of the ligand 2 to ligand 8 transformation in BRD4.}}
	\label{fig:pmf_com_BRD4}
\end{figure}

\begin{figure}[!h]
	\centering
	\includegraphics[width=0.9\textwidth]{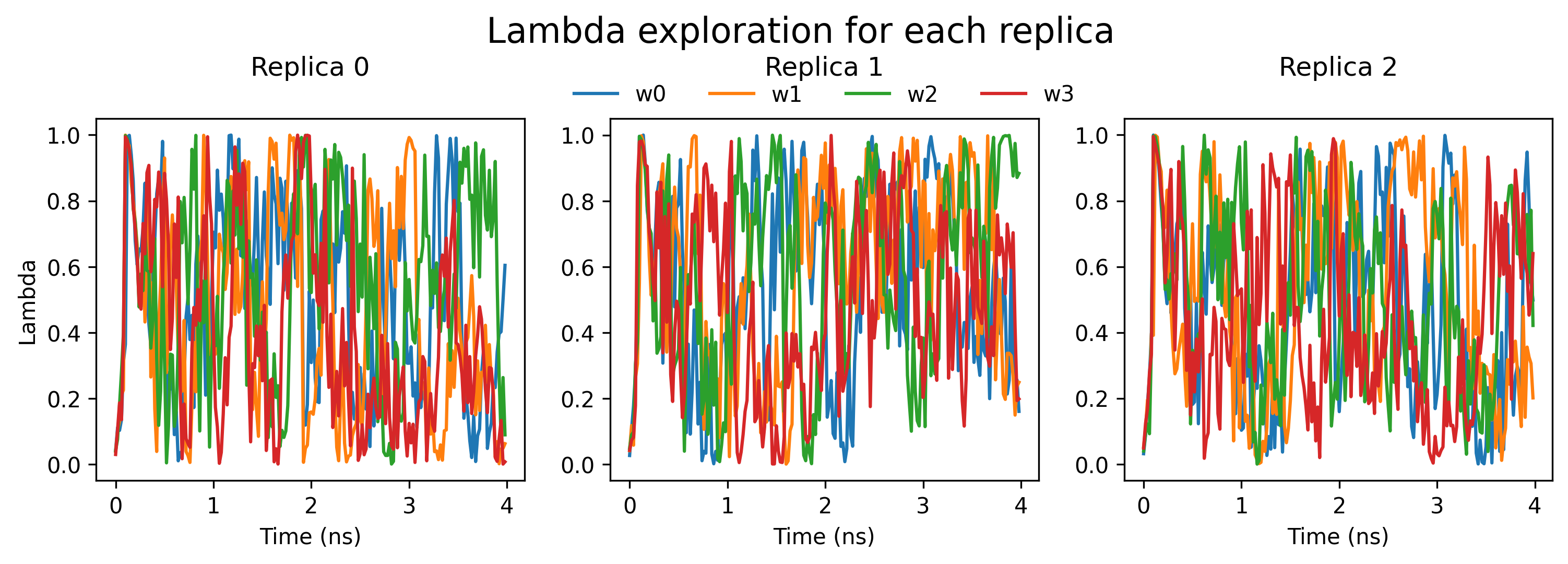}   
\caption{\textbf{Lambda fluctuation over time for the complex phase of the ligand 2 to ligand 8 transformation in BRD4.}}
	\label{fig:lam_com_BRD4}
\end{figure}

\begin{figure}[!h]
	\centering
	\includegraphics[width=0.9\textwidth]{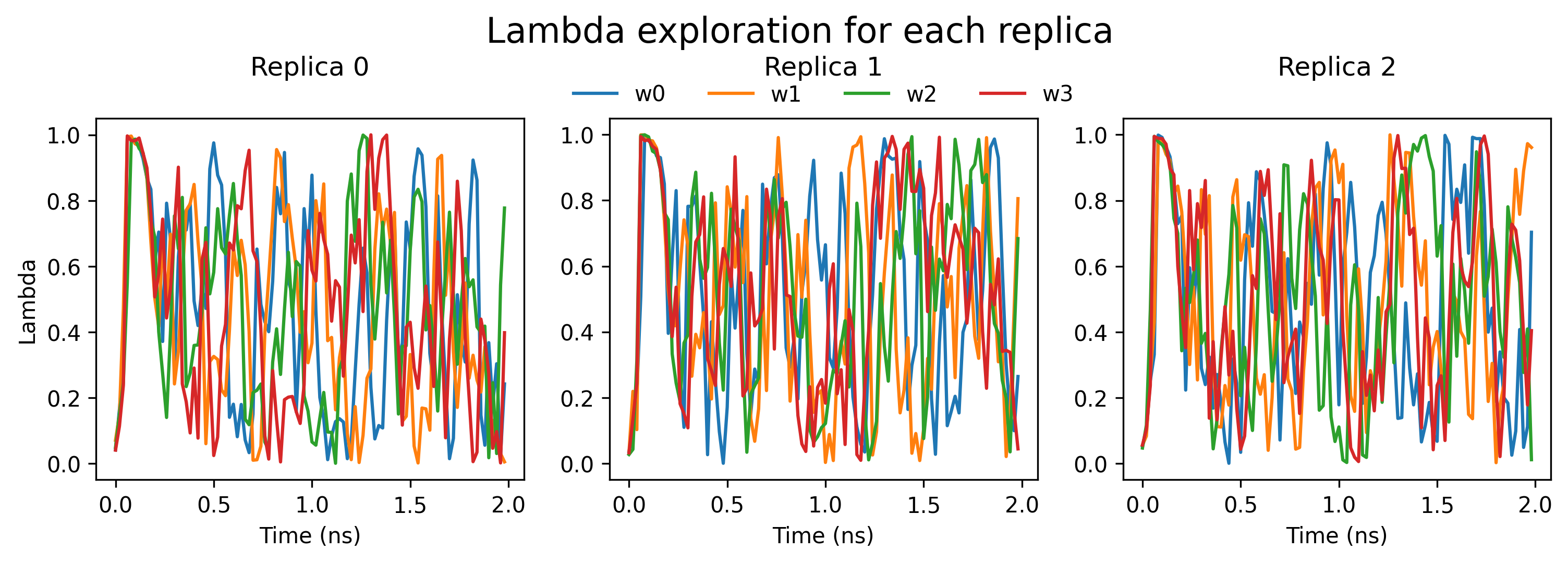}   
\caption{\textbf{Lambda fluctuation over time for the solvent phase of the ligand 10 to ligand 12 transformation.}}
	\label{fig:lam_sol_BRD4}
\end{figure}

\clearpage

\subsection{P38}

\begin{figure}[!h]
	\centering
	\includegraphics[width=0.9\textwidth]{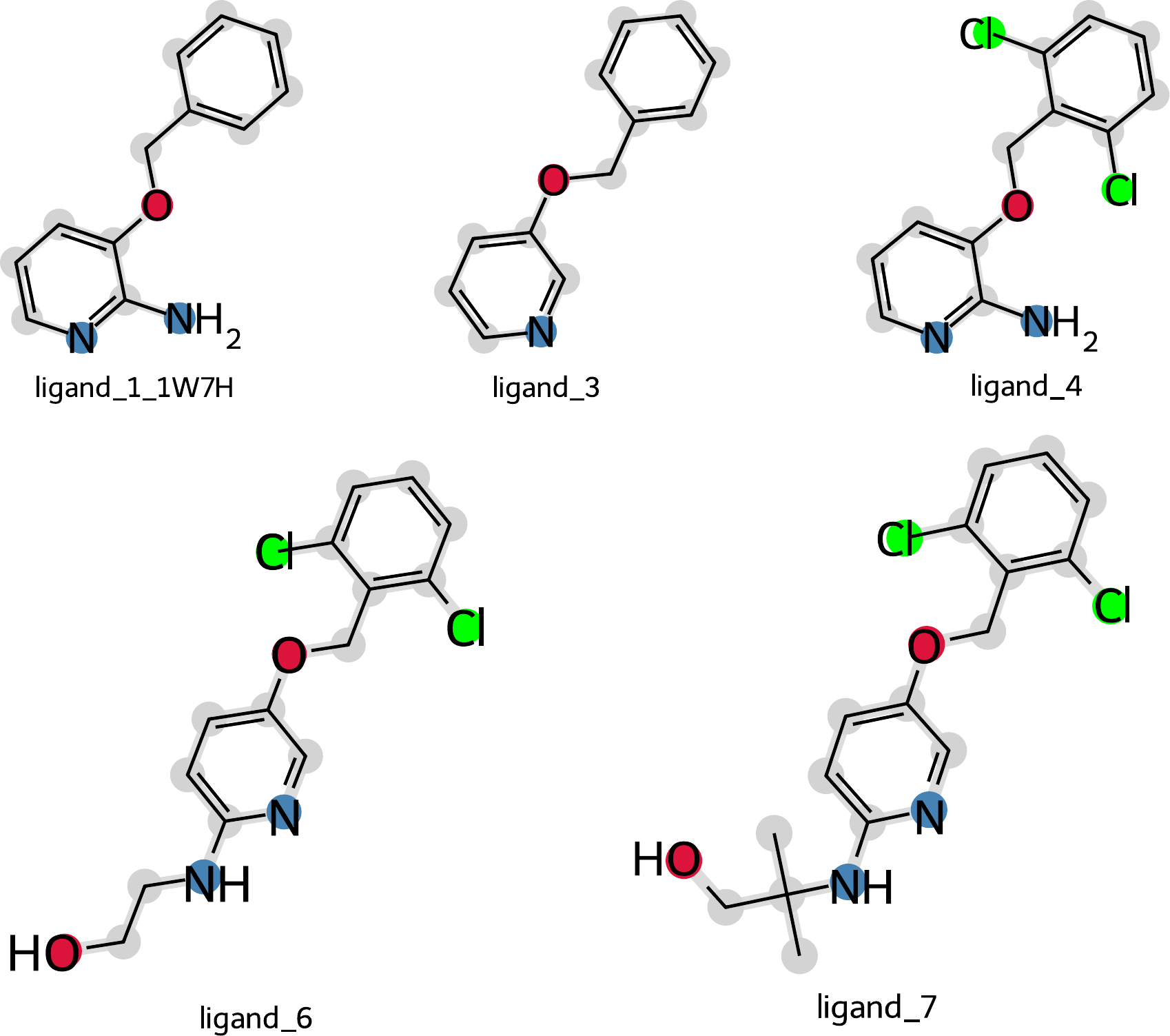}   
\caption{\textbf{2D structures of all ligands complexed with P38.}}
	\label{fig:lig_p38}
\end{figure}

\begin{figure}[!h]
	\centering
	\includegraphics[width=0.3\textwidth]{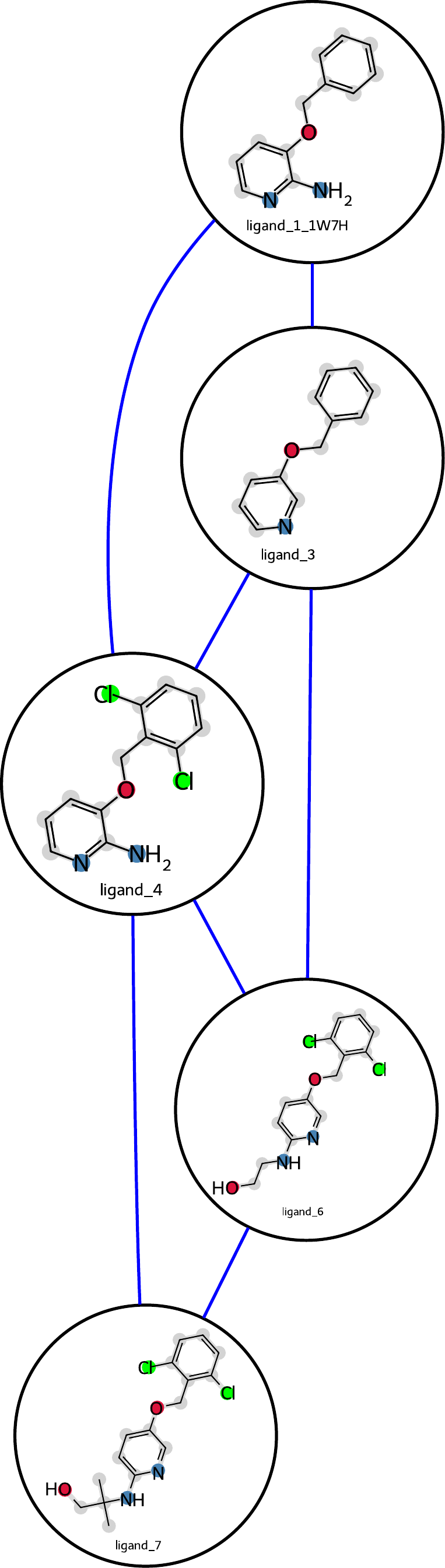}   
\caption{\textbf{LoMap perturbation network for P38 ligand pairs}. Nodes represent individual ligands, and edges indicate the simulated transformation paths.}
	\label{fig:lig_p38_lomap}
\end{figure}


\begin{figure}[!h]
	\centering
	\includegraphics[width=0.6\textwidth]{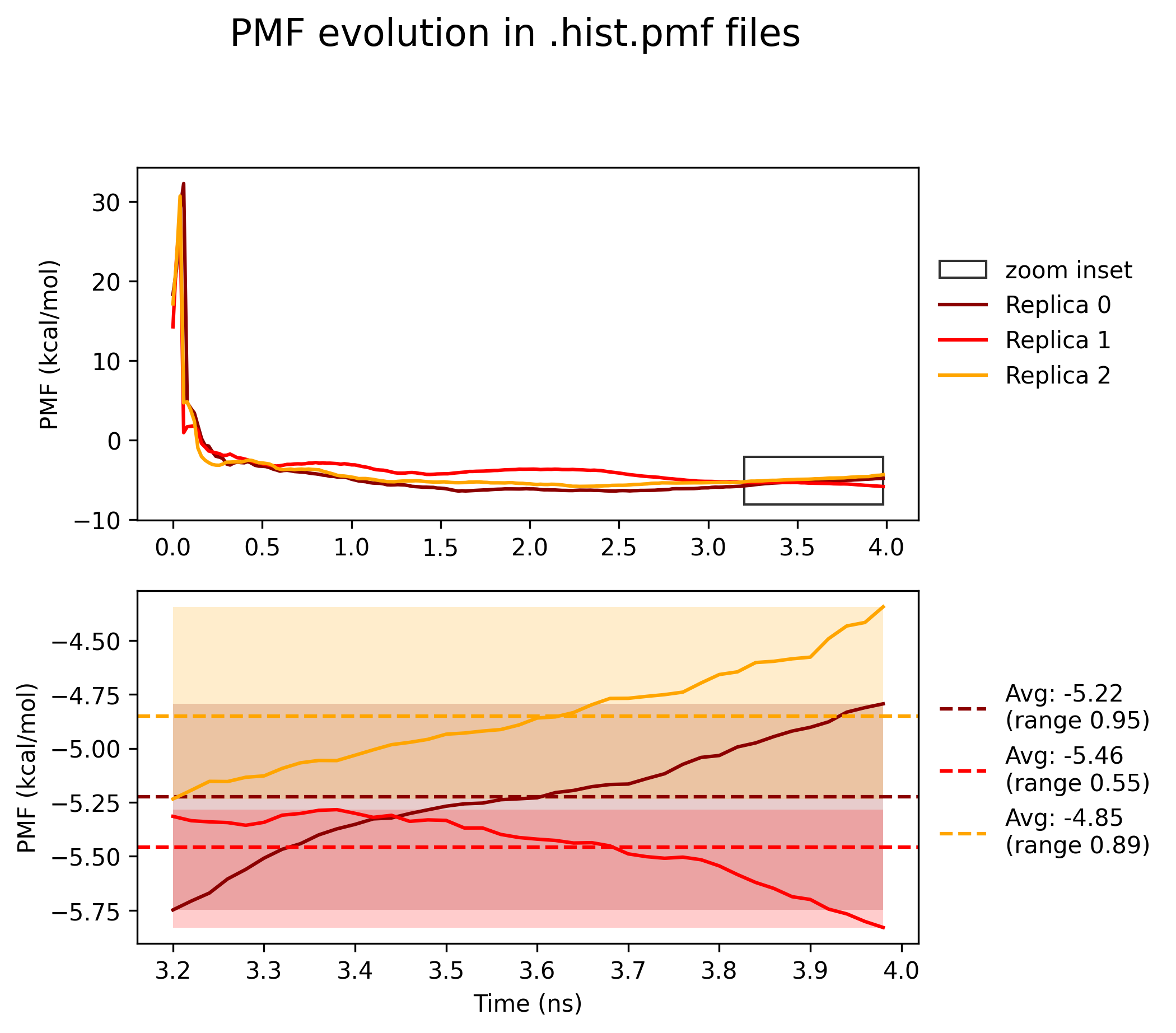}   
\caption{\textbf{Convergence plot of the Potential of Mean Force (PMF) for the complex phase of the ligand 4 to ligand 7 transformation in P38.}}
	\label{fig:pmf_com_P38}
\end{figure}

\begin{figure}[!h]
	\centering
	\includegraphics[width=0.9\textwidth]{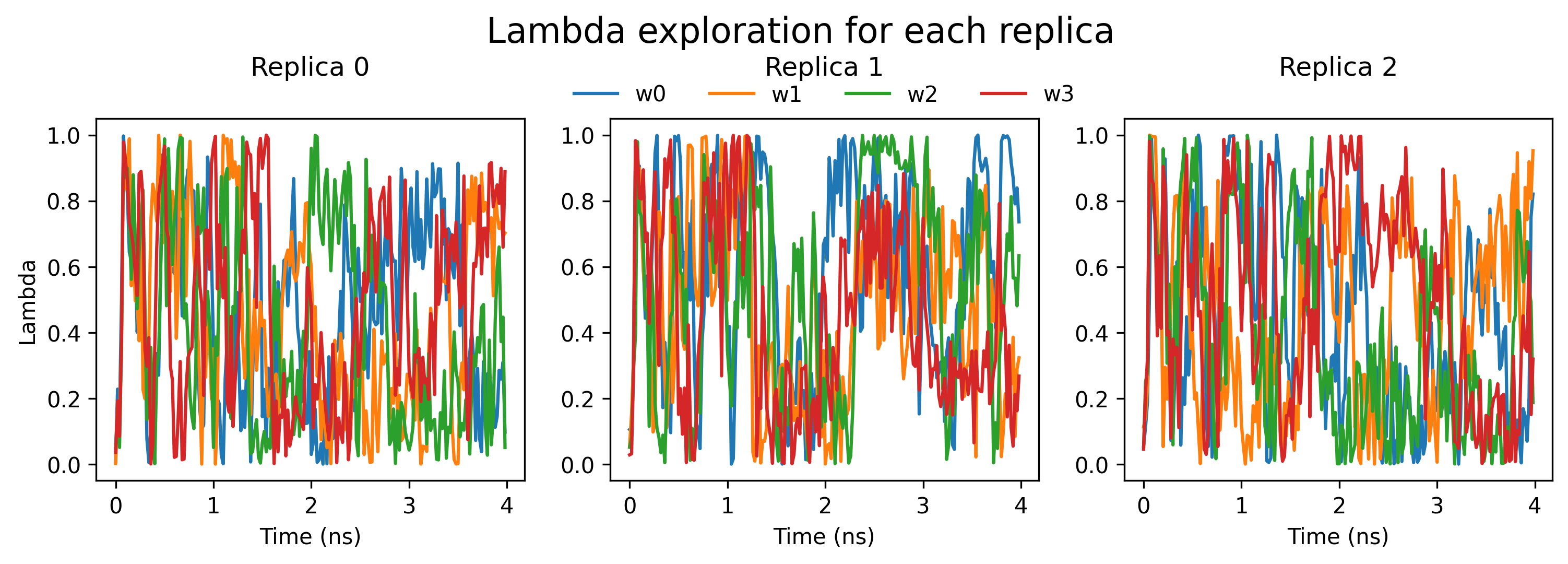}   
\caption{\textbf{Lambda fluctuation over time for the complex phase of the ligand 4 to ligand 7 transformation in P38.}}
	\label{fig:lam_com_P38}
\end{figure}

\begin{figure}[!h]
	\centering
	\includegraphics[width=0.6\textwidth]{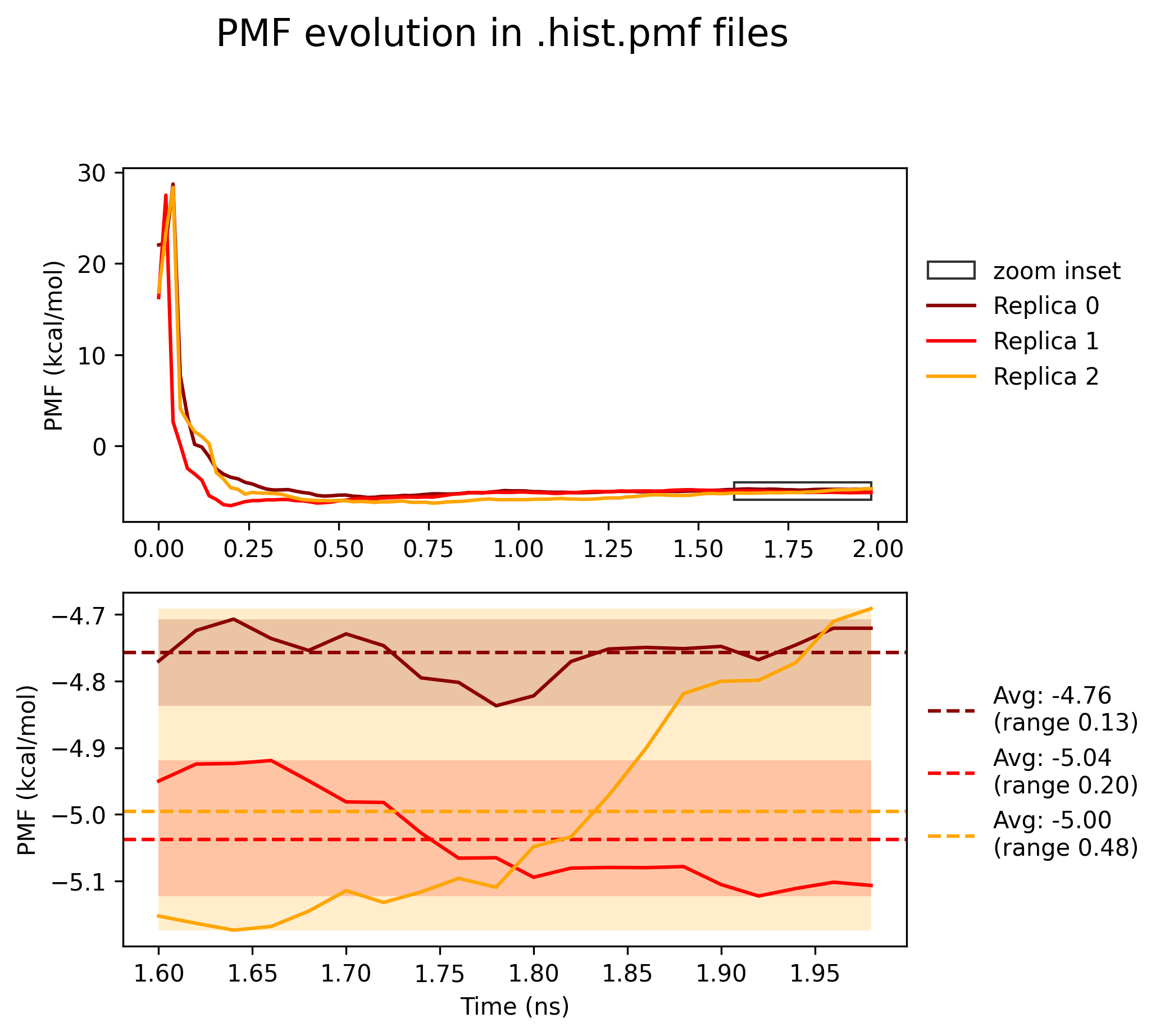}   
\caption{\textbf{Convergence plot of the Potential of Mean Force (PMF) for the solvent phase of the ligand 4 to ligand 7 transformation in P38.}}
	\label{fig:pmf_sol_P38}
\end{figure}

\begin{figure}[!h]
	\centering
	\includegraphics[width=0.9\textwidth]{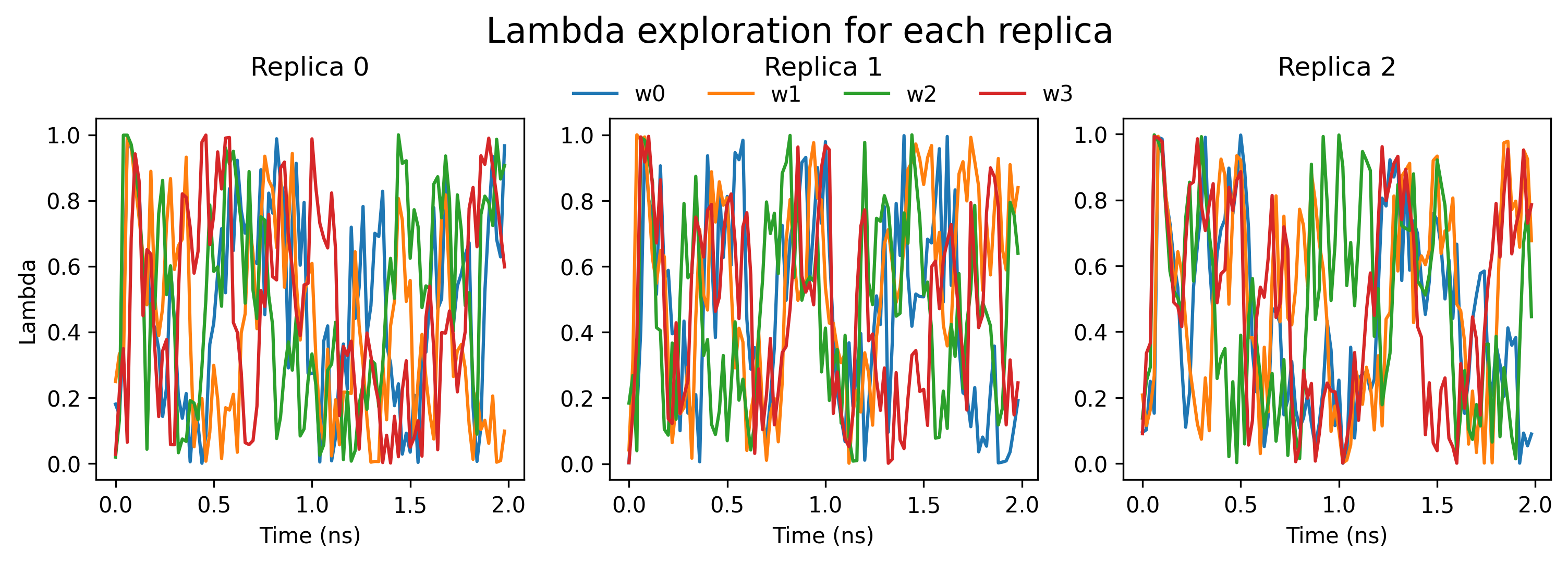}   
\caption{\textbf{Lambda fluctuation over time for the solvent phase of the ligand 4 to ligand 7 transformation in P38.}}
	\label{fig:lam_sol_P38}
\end{figure}

\clearpage

\subsection{CHK1}

\begin{figure}[!h]
	\centering
	\includegraphics[width=0.9\textwidth]{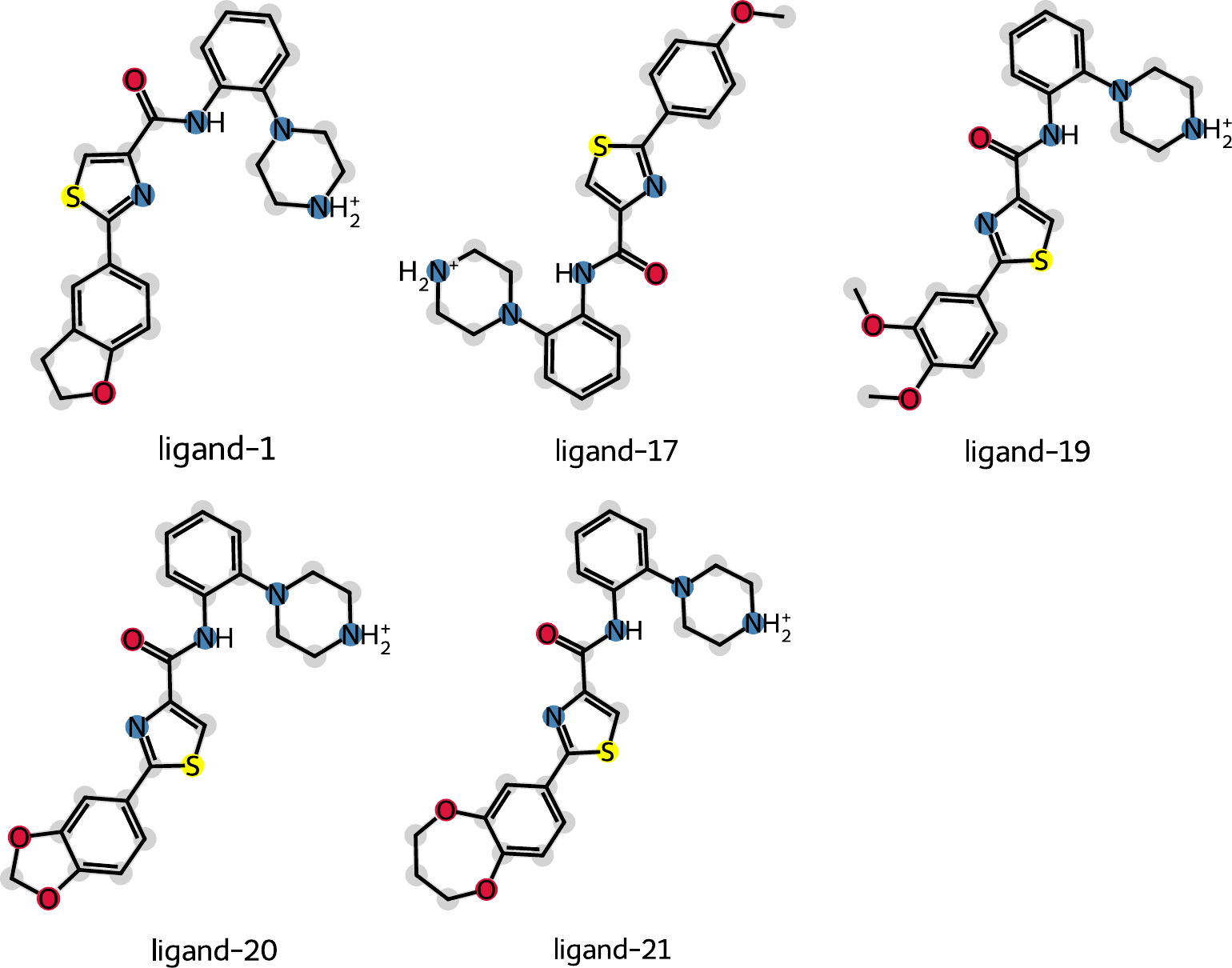}   
     \caption{\textbf{2D structures of all ligands complexed with CHK1.}}
	\label{fig:lig_CHK1}
\end{figure}

\begin{figure}[!h]
	\centering
	\includegraphics[width=0.6\textwidth]{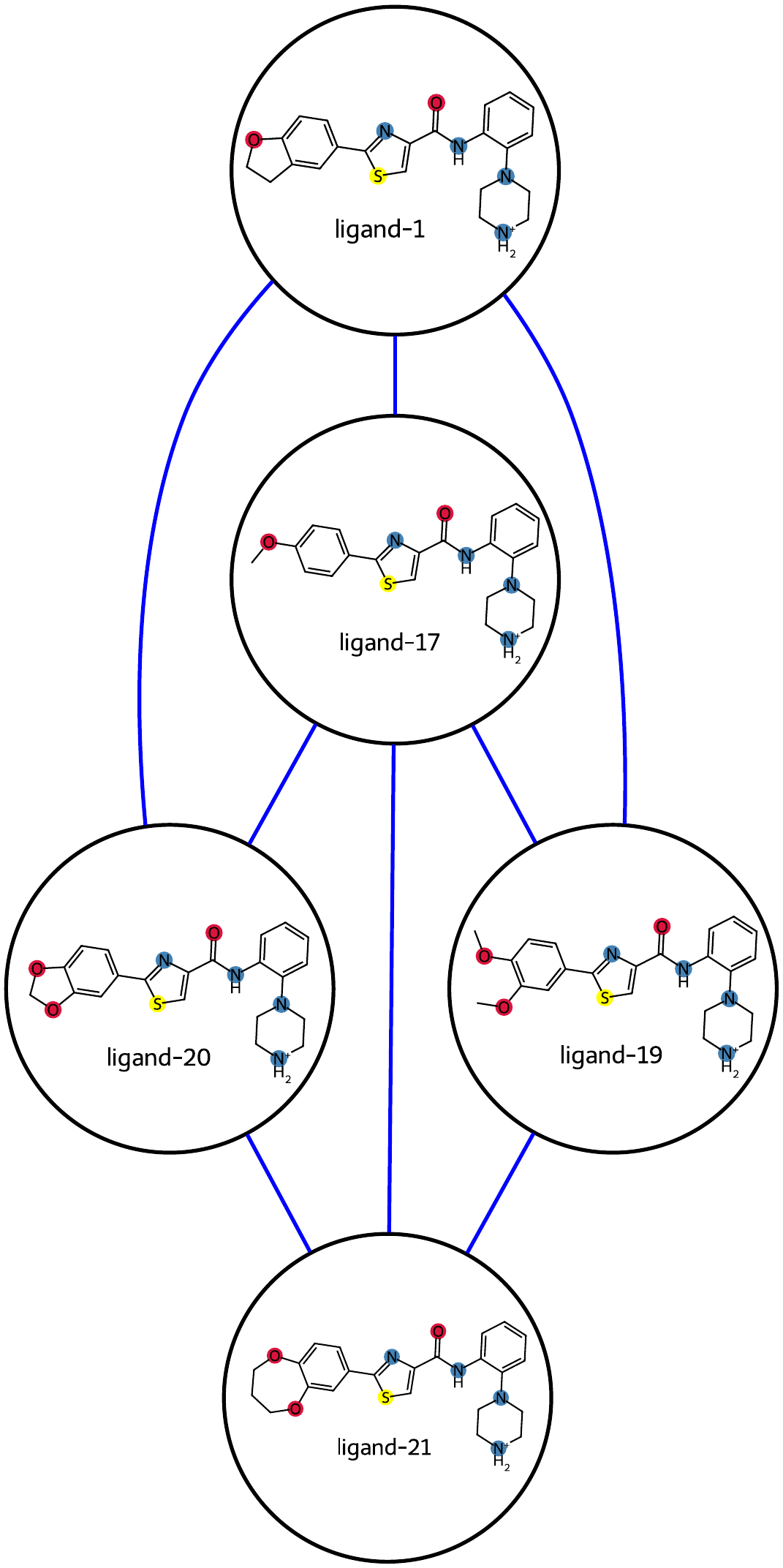}   
\caption{\textbf{LoMap perturbation network for CHK1 ligand pairs}. Nodes represent individual ligands, and edges indicate the simulated transformation paths.}
	\label{fig:LIG_CHK1_LOMAP}
\end{figure}


\begin{figure}[!h]
	\centering
	\includegraphics[width=0.6\textwidth]{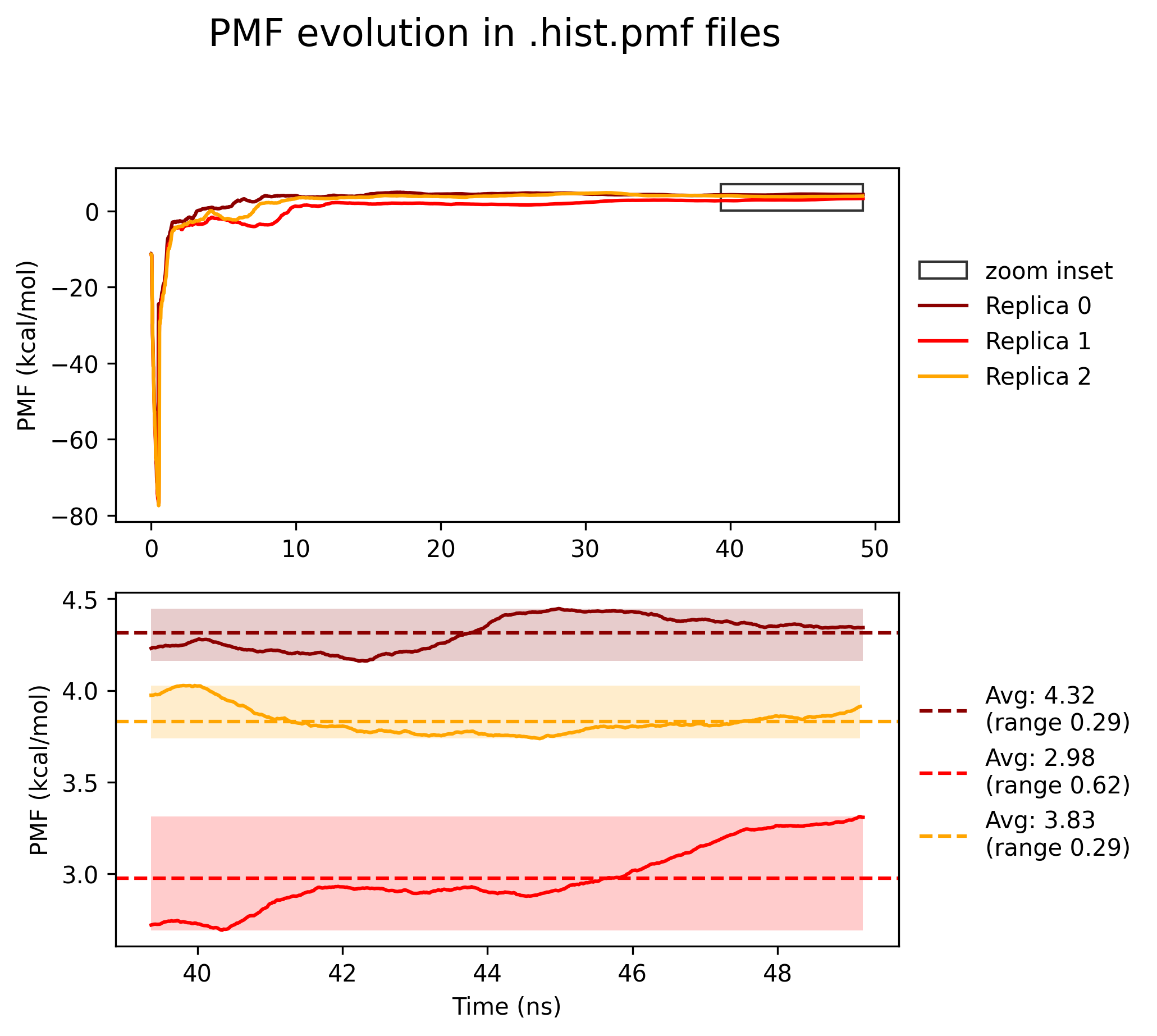}   
\caption{\textbf{Convergence plot of the Potential of Mean Force (PMF) for the complex phase of the ligand 19 to ligand 21 transformation in CHK1.}}
	\label{fig:pmf_com_CHK1}
\end{figure}

\begin{figure}[!h]
	\centering
	\includegraphics[width=0.9\textwidth]{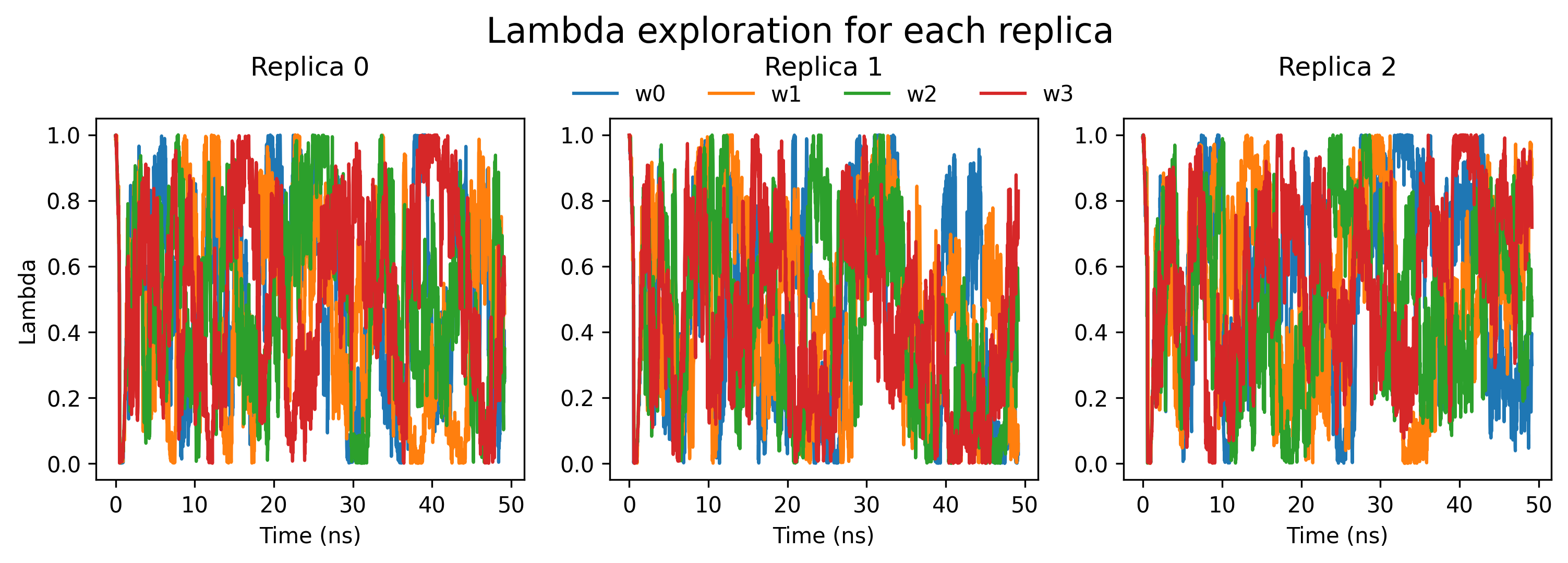}   
\caption{\textbf{Lambda fluctuation over time for the complex phase of the ligand 19 to ligand 21 transformation in CHK1.}}
	\label{fig:lam_com_CHK1}
\end{figure}

\begin{figure}[!h]
	\centering
	\includegraphics[width=0.6\textwidth]{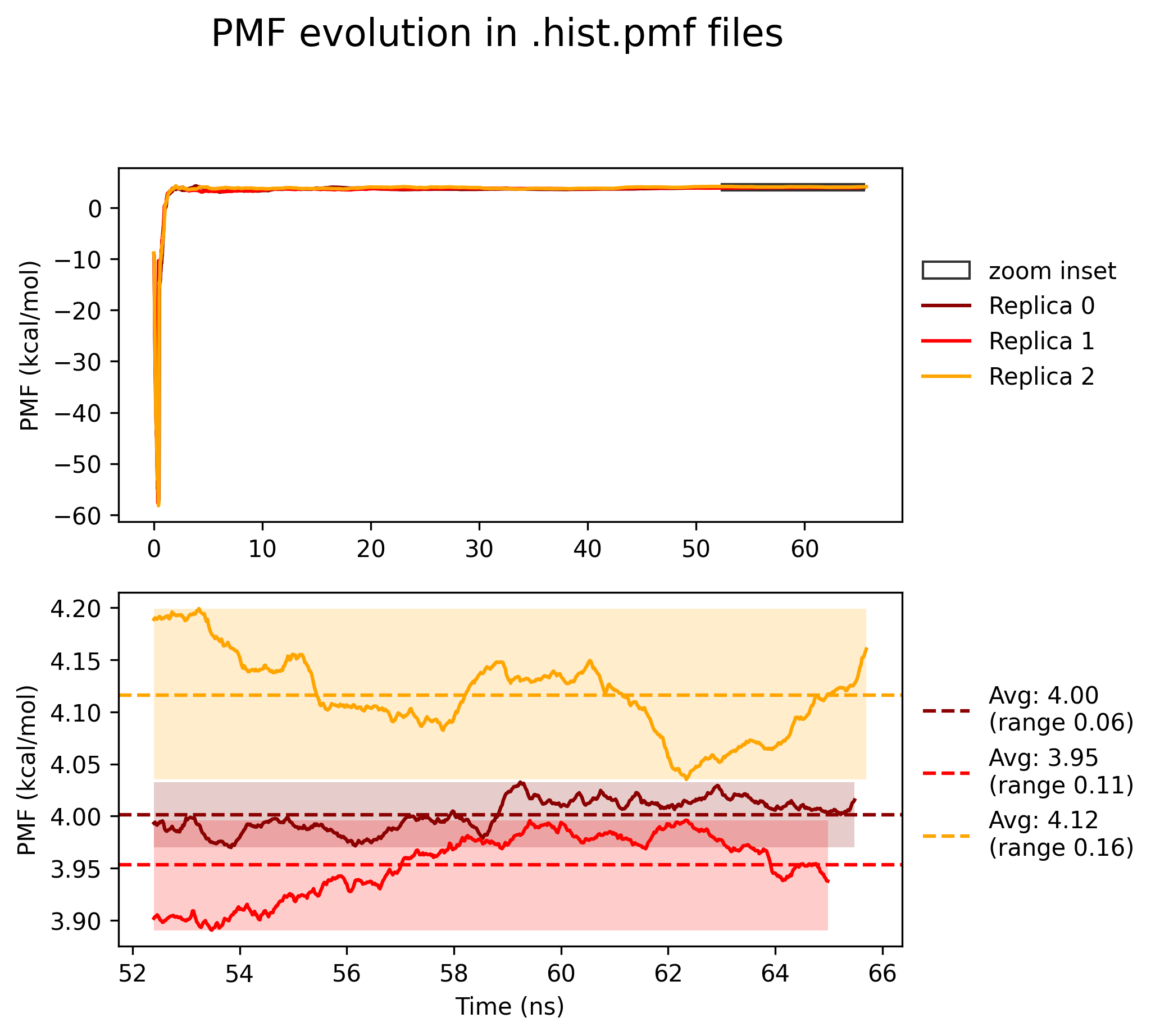}   
\caption{\textbf{Convergence plot of the Potential of Mean Force (PMF) for the solvent phase of the ligand 19 to ligand 21 transformation in CHK1.}}
	\label{fig:pmf_sol_CHK1}
\end{figure}

\begin{figure}[!h]
	\centering
	\includegraphics[width=0.9\textwidth]{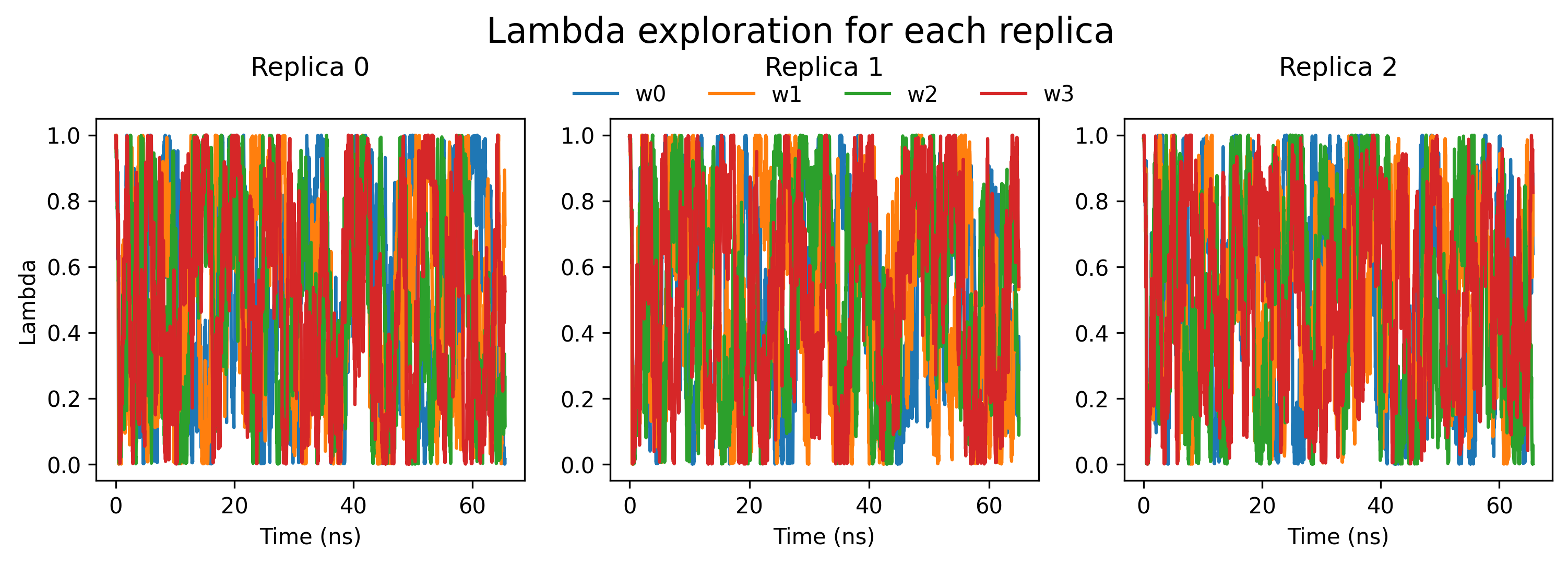}   
\caption{\textbf{Lambda fluctuation over time for the solvent phase of the ligand 19 to ligand 21 transformation in CHK1.}}
	\label{fig:lam_sol_CHK1}
\end{figure}

\clearpage

\subsection{BEN - Multisite simulations}

\begin{figure}[!h]
    \centering
    \includegraphics[width=0.5\linewidth]{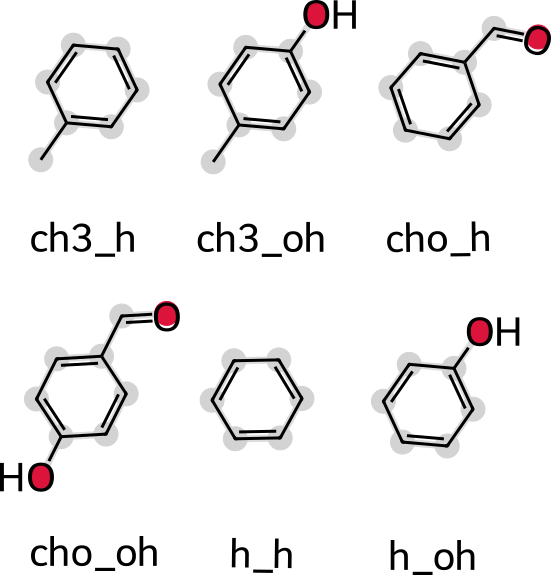}
    \caption{\textbf{2D structures of all benzene derivatives for which Hydration Free Energies were computed.} Ligand names are composed of the substituents carried by the benzene cycle, separated by an underscore symbol ("h\_h" designates benzene, "h\_oh" toluene, \textit{etc.}).}
    \label{fig:lig_BEN}
\end{figure}

\begin{figure}[!h]
    \centering
    \includegraphics[width=0.9\textwidth]{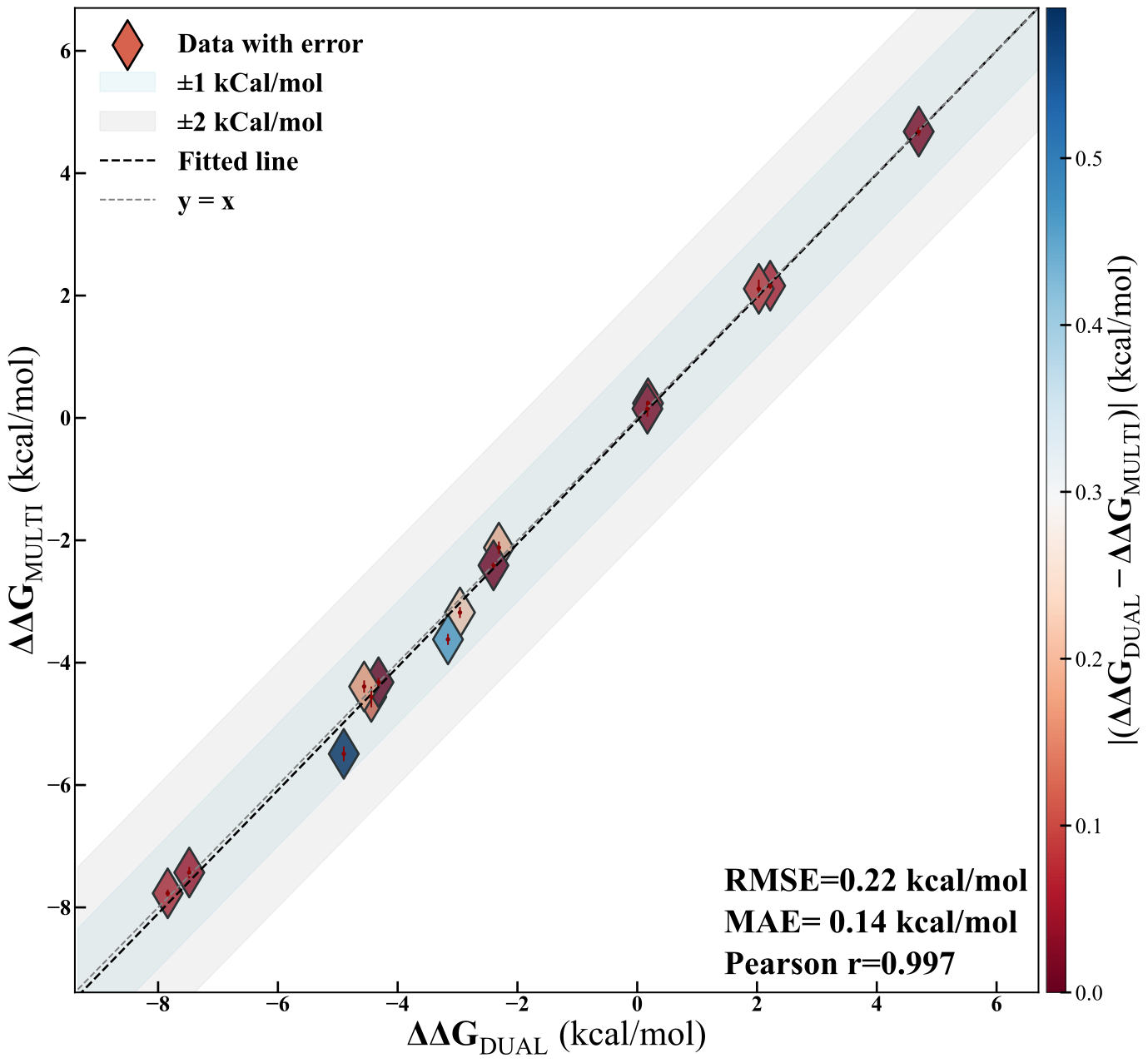}
    \caption{\textbf{Hydration Free Energies computed using the Multiple vs. Dual topology approaches}, on the 15 pairs built from the six molecules presented above. Error bars represent the standard deviation obtained from three independent repeats. Points are colored depending on the absolute error between values obtained with the dual and multiple topology approaches.} 
    \label{fig:HFE_dual_vs_multi}
\end{figure}

\begin{table}[]
\begin{tabular}{llrr}
\hline
\textbf{Ligand 1} & \textbf{Ligand 2} & $\Delta\Delta G_\text{dual} \pm $ std & $\Delta \Delta G_{\text{multi}} \pm $ std  \\ \hline
h\_h               & h\_oh              & -4.44 $\pm$ 0.03 & -4.56 $\pm$ 0.17 \\
h\_h               & ch3\_h             & 0.18  $\pm$ 0.07 & 0.24  $\pm$ 0.02 \\
h\_h               & ch3\_oh            & -4.32 $\pm$ 0.12 & -4.32 $\pm$ 0.07 \\
h\_h               & cho\_h             & -2.31 $\pm$ 0.07 & -2.12 $\pm$ 0.10 \\
h\_h               & cho\_oh            & -7.48 $\pm$ 0.12 & -7.43 $\pm$ 0.11 \\
h\_oh              & ch3\_h             & 4.70  $\pm$ 0.07 & 4.68  $\pm$ 0.07 \\
h\_oh              & ch3\_oh            & 0.17  $\pm$ 0.09 & 0.15  $\pm$ 0.13 \\
h\_oh              & cho\_h             & 2.22  $\pm$ 0.02 & 2.16  $\pm$ 0.05 \\
h\_oh              & cho\_oh            & -2.96 $\pm$ 0.38 & -3.18 $\pm$ 0.36 \\
ch3\_h             & ch3\_oh            & -4.56 $\pm$ 0.09 & -4.39 $\pm$ 0.10 \\
ch3\_h             & cho\_h             & -2.40 $\pm$ 0.09 & -2.41 $\pm$ 0.05 \\
ch3\_h             & cho\_oh            & -7.84 $\pm$ 0.05 & -7.77 $\pm$ 0.32 \\
ch3\_oh            & cho\_h             & 2.03  $\pm$ 0.13 & 2.11  $\pm$ 0.15 \\
ch3\_oh            & cho\_oh            & -3.16 $\pm$ 0.24 & -3.62 $\pm$ 0.31 \\
cho\_h             & cho\_oh            & -4.90 $\pm$ 0.25 & -5.49 $\pm$ 0.21 \\ \hline
\end{tabular}
\caption{\textbf{Relative hydration free energies for a family of benzene-derivated compounds.} Columns "Dual" and "Multi" report the relative hydration free energies calculated using a dual- and multiple-topology scheme, respectively. All quantities reported in kcal.mol\textsuperscript{-1}. "std" denotes the standard deviation computed on three replicas. Ligand names are composed of the substituents carried by the benzene cycle, separated by an underscore symbol ("h\_h" designates benzene, "h\_oh" toluene, \textit{etc.}).}
\end{table}

\begin{figure}
    \centering
    \includegraphics[width=0.9\textwidth]{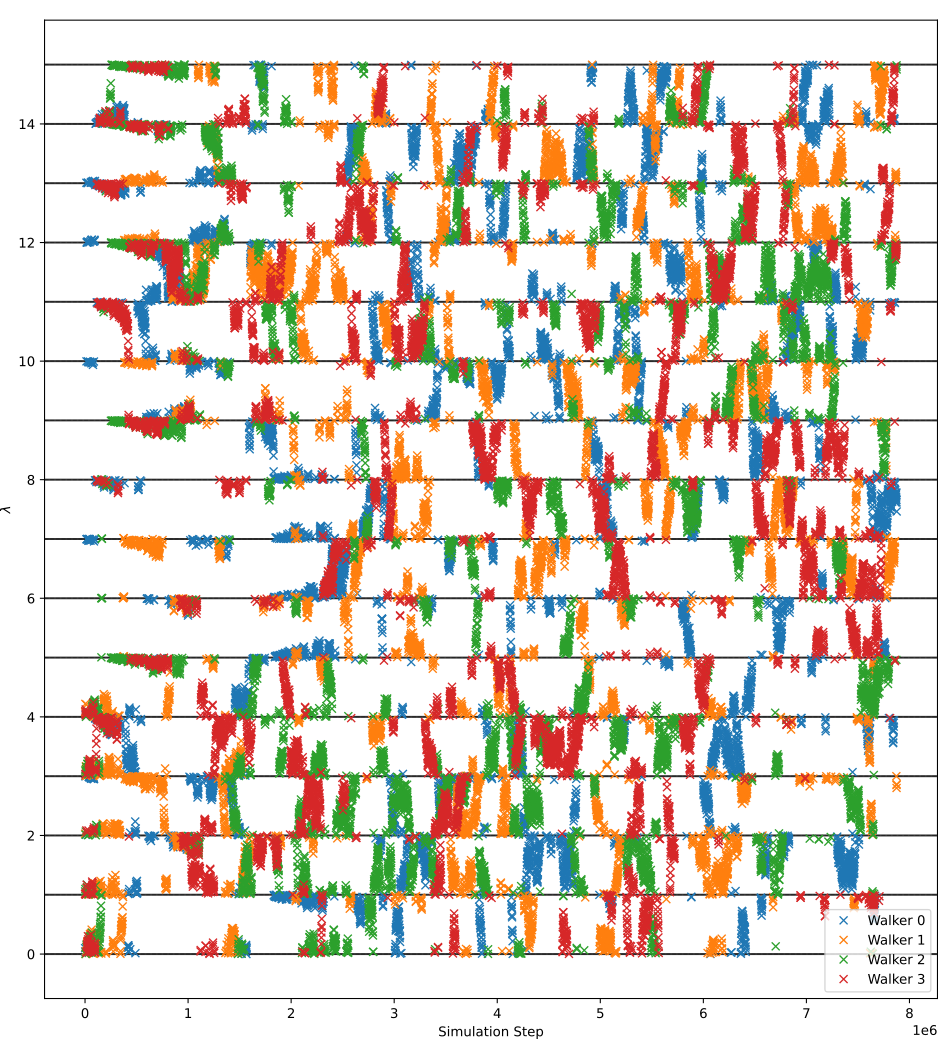}
    \caption{\textbf{Lambda fluctuation over time} for the solvent phase simulation of one replica, in the case of the multiple topology calculation of hydration free energies. Each color designates a different walker. Timestep used was 2~fs. \\
    Each $\lambda$ subdivision of length 1 ($[0;1], [1;2]$, etc.) shows the exploration along an edge of the graph. 
    }
    \label{fig:lam_multisite}
\end{figure}

\clearpage

\clearpage

\section{Supplementary Methods}
\subsection{Alchemical Sampling Parameters and Discretization}

To ensure numerical stability and high resolution along the alchemical path, the following parameters were employed for the $\lambda$ collective variable:

\begin{itemize}
    \item \textbf{ABF Discretization:} The alchemical gradient was accumulated using a fixed bin width of 0.01 (100 bins). A minimum threshold of 10000 samples per bin was required before the full biasing force was applied, ensuring that initial force estimates were sufficiently converged to avoid numerical noise.
    \item \textbf{OPES Adaptive Kernels:} The OPES bias was constructed using an adaptive kernel bandwidth. By dynamically adjusting the bandwidth according to the local sample density, the algorithm suppresses statistical noise in poorly sampled regions while maintaining high precision in well-sampled areas.
    \item \textbf{Update Frequency:} OPES bias was updated with a stride of 300 steps. This frequency ensures that the driving force remains smooth and consistent with the underlying dynamics of the protein-ligand complex.
\end{itemize}

\clearpage

\subsection{Drivers of Protocol Efficiency and Stability}

To isolate the contributions of the individual components of the Dual-LAO protocol, we conducted a study focusing on (i) the enhanced sampling method and (ii) the structural restraints. 

\begin{figure}[!h]
	\centering
	\includegraphics[width=0.9\textwidth]{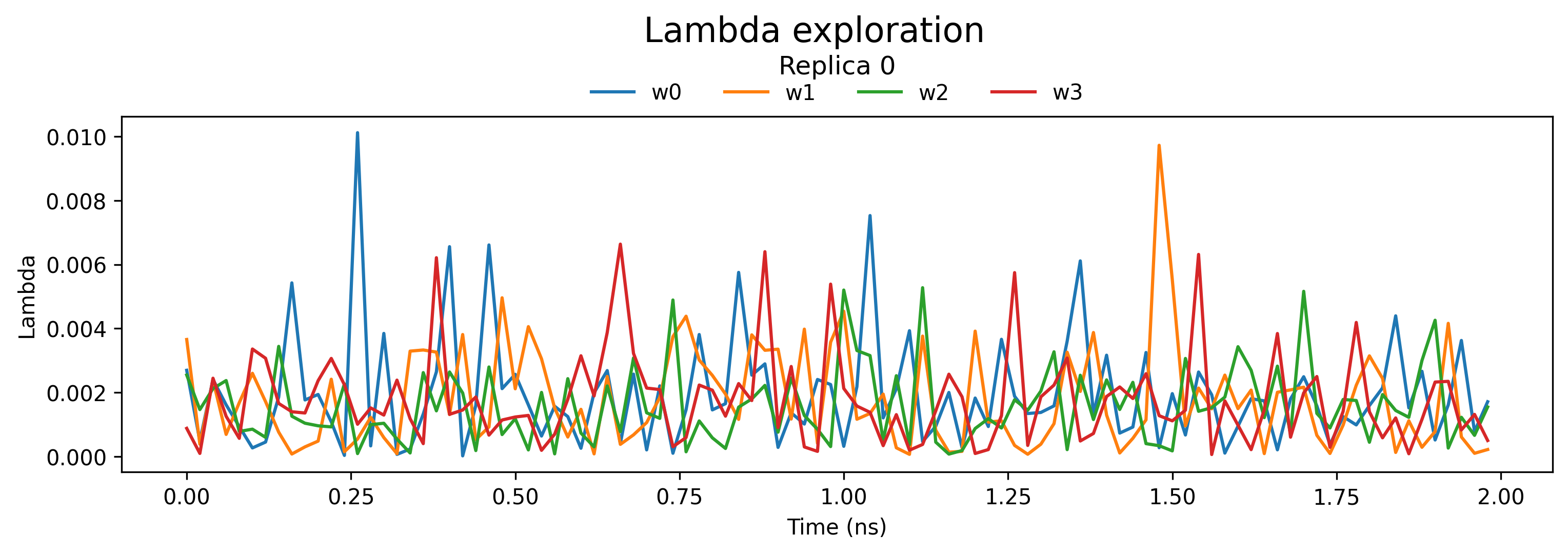}   
\caption{\textbf{Lambda fluctuation over time for the solvent phase of the ligand 10 to ligand 11 transformation in PWWP1 without ABF-OPES.}}
	\label{fig:lam_sol_noOPES}
\end{figure}

The acceleration of alchemical transitions was evaluated in the solvent phase to remove the influence of binding site restraints. As illustrated in Fig.~\ref{fig:lam_sol_noOPES}, without the L-ABF-OPES$_{e}$ method, the $\lambda$ variable fails to explore the alchemical landscape effectively within a 2 ns timescale, remaining trapped near the starting state. In contrast, the inclusion of the enhanced sampling method enables rapid, multiple round-trips across the full alchemical range ($\lambda=0$ to $\lambda=1$). This demonstrates that L-ABF-OPES$_{e}$ is the primary driver of the reported speedup.

\begin{figure}[!h]
	\centering
	\includegraphics[width=0.9\textwidth]{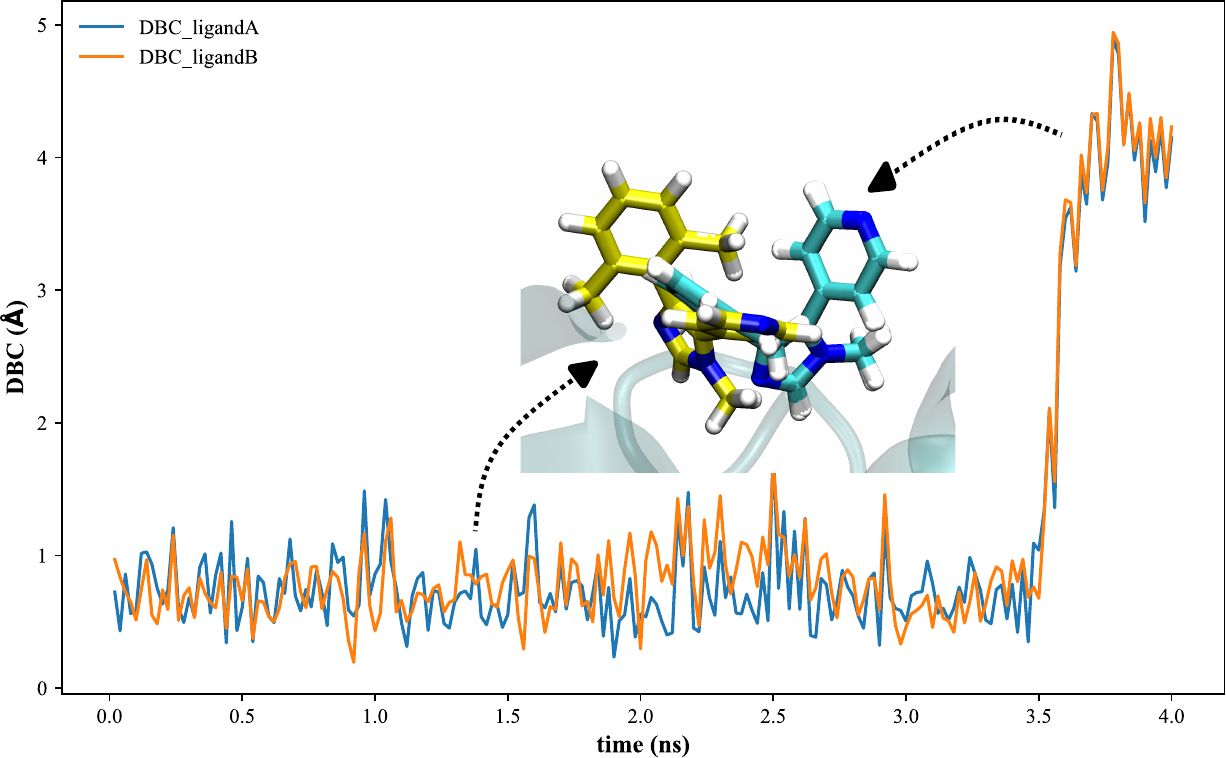}   
\caption{\textbf{Time evolution of the Distance-Based Coordinate (DBC) for the lig10 $\to$ lig11 transformation in the PWWP1 complex phase } without the application of dual-DBC restraints. The corresponding molecular snapshots illustrate that while the ligand in yellow represents the original binding pose, the ligand in cyan shows the drifted configuration.}
	\label{fig:DBC_time}
\end{figure}

The role of the dual-DBC restraints was evaluated in the complex phase using the PWWP1 lig10 $\to$ lig11 system. As shown in Fig.~\ref{fig:DBC_time}, in the absence of these restraints, the ligand in the decoupled or partially decoupled states lacks the necessary intermolecular forces to maintain its orientation within the pocket. This leads to the ligand drifting away from the reference binding pose (highlighted in yellow in the figure). Such detachment prevents the system from maintaining a consistent binding mode, resulting in a failure of thermodynamic closure and physically irrelevant $\Delta\Delta G$ values.

Finally, the dual-topology framework acts as the structural foundation that facilitates these transitions. By providing a smoother potential energy surface compared to single-topology schemes, it prevents numerical instabilities that would occur during bond-breaking events and ensures that the decoupled ligand remains associated with the system. Together, these results confirm that while the enhanced sampling drives the speed of convergence, the dual-DBC and dual-topology components are strictly required to maintain the physical validity of the alchemical transformation.

\subsection{Calculation of ABFE ($\Delta G$) from RBFE ($\Delta\Delta G$)}

The Absolute Binding Free Energies ($\Delta G$) for the series of ligands presented are derived from the calculated Relative Binding Free Energies ($\Delta\Delta G$) using a statistical method called "\text{Weighted Least-Squares Fitting}" of the $\text{RBFE}$ network \cite{wang2015accurate}.

\subsubsection{Weighted Least-Squares Fitting}

The set of calculated $\Delta\Delta G$ values constitutes an \textbf{overdetermined system} of linear equations. The fitting procedure finds the single set of $\Delta G$ values for all ligands ($L_i$) that minimizes the sum of squared residuals ($\text{RMSE}$) across all calculated transformations ($\Delta\Delta G_{i \to j}^{\text{calc}}$):
\[
\text{Minimize} \; \sum_{i \to j} w_{i \to j} \left[ (\Delta G_{L_j} - \Delta G_{L_i}) - \Delta\Delta G_{i \to j}^{\text{calc}} \right]^2
\]

The minimization is \text{weighted} ($w_{i \to j}$) by the inverse variance of the calculated $\Delta\Delta G$ values ($w \propto 1/\sigma^2$). This weighting ensures that the most precise $\Delta\Delta G$ calculations (those with the smallest statistical uncertainty) contribute most significantly to the final fitted $\Delta G$ values.

\subsubsection{Anchoring and Final Results}

The least-squares fit initially yields a relative set of $\Delta G$ values ($\Delta G^{\text{rel}}$). To place these results onto the experimental scale, the entire network is shifted by aligning the $\Delta G^{\text{rel}}$ values to the corresponding experimental $\Delta G^{\text{exp}}$ data for all available ligands. The final reported $\Delta G$ values represent the statistically most probable and thermodynamically consistent binding free energies derived from the calculated $\text{RBFE}$ network, along with the rigorously propagated statistical uncertainty.

\clearpage

\subsection{Computational Cost and Performance Metrics}

\subsubsection{Ligand Parameterization}
Ligand parameters for the polarizable AMOEBA force field were generated using the \texttt{Poltype} 2 package~\cite{poltype2}. The parameterization process, which includes quantum mechanical (QM) calculations for multipoles and polarizability, was performed on CPU-based HPC nodes. Typical runs utilized approximately 60,000 MB (60 GB) of RAM. The wall-clock time for parameterization varied by ligand size but generally ranged from 2 to 4 hours per molecule.

\subsubsection{Simulation Throughput}
The performance of the \text{Tinker-HP} (Version 1.2) engine was evaluated across different generations of NVIDIA GPU architectures using the BAOAB-
RESPA1 integrator ($10$~fs outer time step). Performance metrics for a representative system containing approximately 60K atoms are summarized in Table~\ref{tab:performance}.

\begin{table}[h!]
\centering
\caption{Computational performance of \text{Tinker-HP} for a typical 60k atom system.}
\label{tab:performance}
\begin{tabular}{lc}
\hline
\textbf{GPU Architecture} & \textbf{Performance (ns per day)} \\ \hline
NVIDIA V100               & ~28                            \\
NVIDIA A100               & ~35                            \\
NVIDIA H100               & ~55                            \\ \hline
\end{tabular}
\end{table}

The reported acceleration factor of 15--30 times refers specifically to the \textbf{alchemical production phase} compared to the aggregate sampling time typically required by state-of-the-art FEP+ protocols to achieve comparable convergence. 

While our protocol includes a 20~ns MD equilibration for the selection of Dual-DBC anchor atoms, this is categorized as a \textbf{system equilibration and parameter-derivation phase}. Following established rigorous free energy workflows (e.g., Alibay \textit{et al.}~\cite{alibay2022evaluating}), such a period of unrestrained NPT MD is a standard prerequisite for ensuring the structural integrity of the protein-ligand complex and for defining the physical restraints of the bound state---such as Boresch-style or Dual-DBC parameters---prior to any alchemical transformation.

Notably, while discrete-window methods like FEP+ typically require shorter equilibration bursts per $\lambda$-window, the aggregate ``hidden'' equilibration cost across 12--24 windows often equals or exceeds this 20~ns baseline. Dual-LAO, as a continuous $\lambda$ method, eliminates the need for these redundant window-specific equilibrations. Furthermore, our analysis suggests that the Dual-DBC atom selection is robust even with significantly shorter trajectories (e.g., 2--5~ns). The core efficiency of Dual-LAO lies in its ability to converge the free-energy landscape in 4~ns per walker, representing a substantial reduction in the total number of force evaluations needed to achieve chemical accuracy, particularly for transformations involving high-energy barriers.

\bibliography{General}